\documentclass[sigconf]{acmart}

\usepackage[yyyymmdd]{datetime}

\usepackage{tabularx}
\usepackage{dcolumn} 
\newcolumntype{d}[1]{D{.}{.}{#1}}
\usepackage{multirow}
\usepackage{adjustbox}

\usepackage{subcaption}
\usepackage{adjustbox}

\usepackage{todonotes}
\let\xtodo\todo
\renewcommand{\todo}[1]{\xtodo[inline,color=green!50]{#1}}

\AtBeginDocument{%
  }

\setcopyright{acmlicensed}
\copyrightyear{2025}
\acmYear{2025}
\acmDOI{10.1145/3749385.3749398}

\acmConference[SportsHCI '25]{ACM SportsHCI Annual Conference on Human-Computer Interaction and Sports}
{November 17 - November 19 2025}{Enschede, The Netherlands}

\acmISBN{978-1-4503-XXXX-X/18/06}

\begin{document}

\title[At the Speed of the Heart]{At the Speed of the Heart:\\ Evaluating Physiologically-Adaptive Visualizations for Supporting Engagement in Biking Exergaming in Virtual Reality}

\author{Oliver Hein}
 \orcid{0009-0009-9962-0165}
\affiliation{
  \institution{University of the Bundeswehr Munich}
  \city{Munich}
  \country{Germany}
}
\email{oliver.hein@unibw.de}

\author{Sandra Wackerl}
 \orcid{0009-0009-3243-5213}
\affiliation{
  \institution{LMU Munich}
  \city{Munich}
  \country{Germany}
}
\email{Sandra.Wackerl@campus.lmu.de}

\author{Changkun Ou}
\orcid{0000-0002-4595-7485}
\affiliation{
  \institution{LMU Munich}
  \city{Munich}
  \country{Germany}
}
\email{research@changkun.de}

\author{Florian Alt}
\orcid{0000-0001-8354-2195}
\affiliation{
  \institution{LMU Munich}
  \city{Munich}
  \country{Germany}
}
\email{florian.alt@lmu.de}

\author{Francesco Chiossi}
 \orcid{0000-0003-2987-7634}
\affiliation{
  \institution{LMU Munich}
  \city{Munich}
  \country{Germany}
}
\email{francesco.chiossi@lmu.de}

\renewcommand{\shortauthors}{Hein et al.}

\vspace{-2mm}\begin{abstract}
Many exergames face challenges in keeping users within safe and effective intensity levels during exercise. Meanwhile, although wearable devices continuously collect physiological data, this information is seldom leveraged for real-time adaptation or to encourage user reflection.
We designed and evaluated a VR cycling simulator that dynamically adapts based on users' heart rate zones. First, we conducted a user study ($N=50$) comparing eight visualization designs to enhance engagement and exertion control, finding that gamified elements like non-player characters (NPCs) were promising for feedback delivery. Based on these findings, we implemented a physiology-adaptive exergame that adjusts visual feedback to keep users within their target heart rate zones.
A lab study ($N=18$) showed that our system has potential to help users maintain their target heart rate zones. Subjective ratings of exertion, enjoyment, and motivation remained largely unchanged between conditions.
Our findings suggest that real-time physiological adaptation through NPC visualizations can improve workout regulation in exergaming. 

\end{abstract}

\vspace{-10mm}\ccsdesc[500]{Human-centered computing~Human computer interaction (HCI)}

\vspace{-10mm}\keywords{Exergaming, Virtual Reality, Physiological Computing, Adaptive Systems, ECG}

\begin{teaserfigure}
\vspace{-4mm}
 \includegraphics[width=\linewidth]{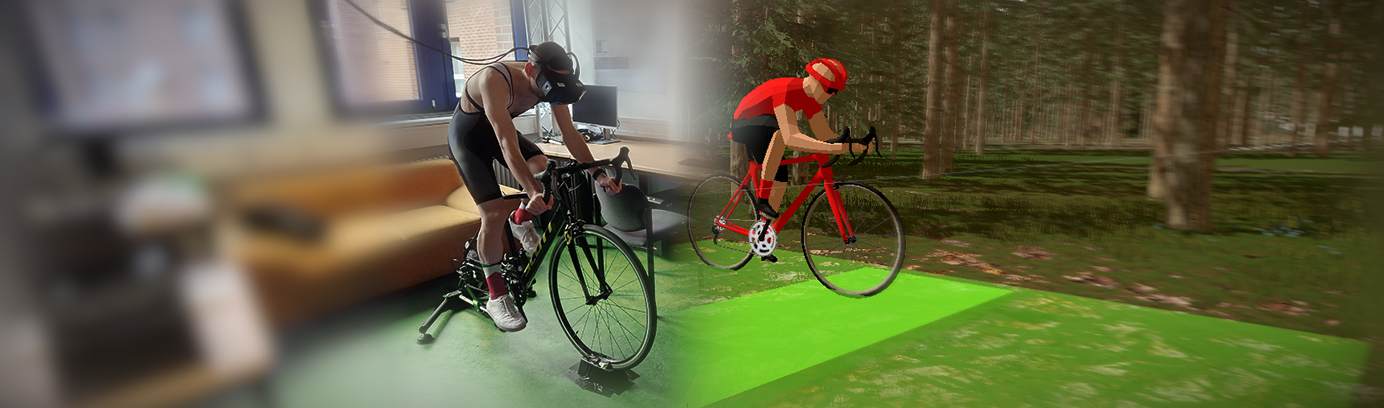}
\vspace{-8mm}
 \caption{
We enhance exergame performance by evaluating eight visual designs, identifying gamified elements such as Non-Playable Characters (NPCs) as most effective. Based on this, we developed a VR cycling simulator featuring an adaptive NPC that adjusts to the user’s heart rate to help maintain ideal cardio levels. The image shows a participant from our second study (left) and the VR environment (right), where the participant cycles through a forest alongside the adaptive NPC, which represents their optimal heart rate.}
 \Description{Two cyclists sitting on a bike. While the left half of the image is real and shows a real participant, the right half shows the virtual scenario we implemented, including a NPC.}
 \label{fig:teaser}
\end{teaserfigure}

\maketitle

\section{Introduction}

Monitoring physiological data, such as heart rate (HR), is widely used in athletic training to optimize performance, prevent overexertion, and enhance endurance. Proper regulation of HR zones during exercise is critical for cardiovascular adaptations, reducing fatigue, and minimizing injury risk \cite{mj1957effects, tanner2012physiological}. While professional athletes leverage HR-based training to fine-tune intensity and recovery, non-professional users often struggle to interpret raw physiological data, leading to ineffective workouts or unsafe exertion levels \cite{Nguyen2023Training, Zhao2021The}. Although modern wearables provide real-time HR tracking, there is a lack of actionable feedback to encourage
user reflection,  making users guess how to adjust their exertion \cite{Tomes2019Ability, Nguyen2023Training}. Bridging the gap between sophisticated physiological tracking and meaningful exercise guidance remains a core challenge.

Digital platforms such as Strava have transformed cycling into a social and competitive sport, while smart trainers like Wahoo\footnote{\url{https://eu.wahoofitness.com/devices/indoor-cycling/bike-trainers}, last accessed \today.}, Garmin\footnote{\url{https://www.garmin.com/en-US/c/sports-fitness/indoor-trainers/}, last accessed \today.}, and Zwift\footnote{\url{https://zwift.com/collections/smart-trainers}, last accessed \today.} enable structured home workouts \cite{xu2021psychological}. Despite these advancements, existing tools still fail to provide real-time physiological adaptation, particularly for users lacking professional coaching or access to diagnostic assessments \cite{Zemkova2018Sport}. Prior research shows that real-time HR feedback can improve training outcomes \cite{li2016wearable}, and social elements, such as virtual training partners, can sustain motivation \cite{agharazidermani2023exploring}. However,  many exergames lack physiological monitoring and adaptive feedback, making it difficult for users to effectively regulate their exertion levels \cite{chiossi2022adapting}. Existing VR exergames, though used in specific domains such as kinematic analysis \cite{munoz2019kinematically} or police training \cite{munoz2020psychophysiological}, rarely integrate physiological signals for real-time adaptive feedback. While early studies suggest that individualized feedback can enhance user experience and performance \cite{sessa2018sports,andres2020introducing}, a gap remains in ensuring exercise is both engaging and physiologically effective  \cite{gao2014field, hochsmann2016effects, kojic2019impact}. 

To address this gap, we investigate how real-time physiological data can enhance VR-based exergaming by dynamically adjusting visual feedback to help users maintain target HR zones. We developed a \emph{VR Cycling Simulator} that adapts in-game elements based on HR changes, allowing users to train at optimal exertion levels (\autoref{fig:teaser}). Our work extends previous research by exploring how different HR visualizations impact exertion control and implementing a closed-loop physiological adaptation system in a VR exergame.

To evaluate this approach, we conducted two studies. In the first study ($N=50$), we examined user preferences for different HR zone visualizations, identifying gamified feedback mechanisms (e.g., NPCs)  as particularly promising for engagement and exertion control. Based on these findings, we implemented an adaptive system that dynamically responded to participants' HR. In a second study ($N=18$), we compared this adaptive system to a random NPC condition, which lacked adaptation and a  baseline control without additional feedback.

Our results show that adaptive feedback via an NPC visualization significantly improved users' ability to maintain target HR zones, though subjective ratings of motivation and exertion remained unchanged. These findings highlight the potential of real-time physiological adaptation for precise, engaging, and safer VR exergame training while also identifying opportunities to further enhance motivation.

By integrating real-time HR-based adaptation, VR exergames can improve training accuracy, engagement, and accessibility  \cite{sessa2018sports, zahabi2020adaptive}. This personalization can benefit athletes of varying fitness levels, expanding the potential impact of exergaming for fitness, rehabilitation, and adaptive health interventions.
However, this study does not seek to demonstrate long-term health impacts or practical deployment but rather explores the feasibility and immediate effects of adaptive HR feedback in a controlled, VR-based cycling environment.

\vspace{1mm}\noindent\textbf{Contribution Statement.} We contribute to the field of physiological computing and adaptive exergaming by
(1) introducing a physiology-adaptive NPC that dynamically adjusts in real-time to optimize controlled exertion rather than merely increasing challenge or competition, as seen in prior ghost AI and opponent-based exergaming; (2) demonstrating the effectiveness of real-time adaptive feedback for maintaining target heart rate zones in VR cycling, distinguishing our work from traditional dynamic difficulty adjustment (DDA), which primarily focuses on task difficulty rather than physiological state regulation; and (3) providing open-source tools and data to advance research in adaptive fitness, physiological computing, and Mixed Reality exergaming. Our findings highlight how personalized, physiologically adaptive systems that prioritize safe, effective, and sustainable training broaden the impact of MR-based exercise applications beyond performance enhancement alone.

\section{Related Work}

In the following, we give an overview of physiologically adaptive systems in VR and summarize the VR exergaming applications and how visualizations are designed to support engagement during sports activities.

\subsection{Physiologically Adaptive Systems in VR}

Physiological computing leverages real-time bodily signals to adapt interactive systems according to a user’s current state \cite{fairclough2017Physiological}. \emph{Virtual Reality} (VR) is particularly conducive to such adaptations due to its immersive nature, which allows precise control over the environments and stimuli presented \cite{chiossi2022adapting}. VR systems can dynamically enhance engagement and well-being by monitoring physiological changes, creating a direct connection between a user’s experience and their bodily reactions \cite{higuera2017psychological, chiossi2025designing}.

Among physiological signals, \emph{Electrocardiography} (ECG) is particularly relevant for VR exergaming. It provides insights into stress, arousal, and physical exertion \cite{biddiss2010active, graves2010physiological}. By tracking HR, systems can dynamically tailor activities to an individual’s cardiovascular state, ensuring users remain within optimal exertion levels \cite{munoz2018closing}. This is crucial in exergames, where ECG offers strong construct validity when examining motivation and performance \cite{fairclough2017Physiological}.

While the \emph{Motivation Intensity Model} (MIM) \cite{richter2016three, richter2014mood} links motivation to perceived difficulty rather than directly to HR, HR can serve as an indirect measure of exertion, which in turn influences perceived effort. In physically demanding tasks, effort mobilization depends on both the perceived challenge and the physiological strain required to meet it. Because HR reflects both physical exertion and autonomic arousal \cite{beh1990achievement, umer2022heart, mosley2022scoping}, it can be leveraged in \emph{closed-loop applications} where task difficulty dynamically adapts to the user's physiological state, maintaining an optimal balance between engagement and effort regulation.

Despite the wealth of sensing options, user-facing representations of physiological signals often remain simplistic. To bridge this gap, \citet{wagener2023weather} showcased a VR system using weather metaphors to visualize stress data, prompting users to reflect on their feelings. Building on these insights, the next section explores how closed-loop physiological feedback can benefit VR exergaming, enhancing training quality and safety by adapting to real-time exertion levels.

\subsection{Closed-Loop Exergaming in VR}
Exergaming is defined a blend of ``exercise'' and ``gaming'',describes video games requiring physical activity as part of the interaction, leveraging movement and physiological signals to create engaging fitness experiences \cite{mueller2018experiencing, mueller2020towards}. Over two decades, exergaming research has explored its potential to improve health, motivation, and user experience \cite{huber2021dynamic, potts2024sweating}.

Exergames often integrate \emph{real-time physiological sensing} (e.g., HR, electrodermal activity, or respiration) to personalize workouts, adapting to an individual’s \emph{fitness level and exertion capacity} \cite{sween2014role, oh2010defining}. This integration spans across various physical activities, from \emph{running and cycling} to \emph{rehabilitation therapy}, where \emph{adaptive difficulty adjustment} ensures users train within optimal intensity ranges \cite{huber2021dynamic, peng2011playing}. Exergames also vary in \emph{immersiveness}, ranging from screen-based motion games (e.g., Kinect, Wii Fit) to \emph{fully immersive VR-based exergames} simulating dynamic, interactive environments \cite{chiossi2024optimizing, chiossi2023adapting}.

VR enhances exergaming by immersing users in visually and aurally rich environments, potentially increasing \emph{motivation and engagement} \cite{chiossi2025adaptive, floyd2021limited}. However, many VR exergames fail to ensure users train at \emph{physiologically appropriate levels}, potentially limiting \emph{fitness benefits and safety} \cite{biddiss2010active, graves2010physiological}. For instance, \emph{non-adaptive exergames} may push users into overly strenuous activity or fail to provide sufficient exertion, making their effectiveness unpredictable \cite{peng2011playing}.
As noted by \citet{martin2021comparing}, numerous researchers have demonstrated that HR-based approaches are both feasible and effective for balancing player abilities and game challenges in exergames. However, these methods are also subject to criticism, as heart rate is an individual metric influenced by a range of internal and external factors.

\emph{Closed-loop exergaming} aims to address this challenge by incorporating \emph{biocybernetic adaptation}, where \emph{real-time physiological signals} (e.g., heart rate zones) are fed back into \emph{game mechanics and difficulty adjustments} \cite{ mueller2018experiencing}. Such \emph{adaptive systems} can dynamically \emph{modulate resistance, speed, or competition intensity} to \emph{maintain target exertion levels}, creating \emph{safer and more effective} workouts \cite{potts2024sweating, huber2021dynamic}. Research shows that \emph{adaptive difficulty control} in exergames can enhance both \emph{engagement and performance}, leading to \emph{long-term adherence to exercise regimens} \cite{schmidt2018impact, wunsche2021rift}.

In this work, we contribute to \emph{physiologically adaptive exergaming} by designing and implementing \emph{adaptive visualizations in VR} to support \emph{engagement and performance}. Our study investigates how \emph{different representations of heart rate zones} influence \emph{user experience, motivation, and exertion control}, providing insights into the \emph{potential of closed-loop VR exergaming for fitness and training applications}.

\subsection{Research on Cycling Simulators}

Recent work on cycling simulators has targeted multiple dimensions, i.e., competition, feedback strategies, realism, comfort, and emotional factors, to advance both user performance and engagement in VR environments \cite{matviienko2024grand}. Given its strong influence on user motivation, a key focus has been \emph{competition}. For example, \citet{shaw2016competition} discovered that virtual ghost opponents or solo play often lead to better performance and enjoyment than cooperative scenarios, suggesting that competitive tension can be a potent motivator. Expanding on this theme, \citet{barathi2018interactive} examined \emph{interactive feedforward} methods, enabling cyclists to race against their own prior performances. This approach improved results over merely racing a generic opponent by preserving intrinsic motivation and personalizing the competitive benchmark. Likewise, \citet{wunsche2021rift} demonstrated that dynamically adjusting an opponent’s pace to users’ heart rates can enhance engagement among less-fit participants, though such balancing risks demotivating highly fit users.

Beyond competition, \emph{realism} is another critical goal for cycling simulators. \citet{schramka2017development} integrated consumer VR devices and motion data to increase immersion, while \citet{michael2020race} introduced ghost NPCs reflecting riders’ historical performances, a tactic shown to enhance both performance and motivation by reinforcing personal feedback loops. To mitigate simulator sickness and further amplify realism, \citet{wintersberger2022development} proposed a subtle tilting mechanism, demonstrating that motion cues can significantly improve cycling experiences without negatively impacting performance or comfort. \citet{wu2023smarcypad} added to these advancements by creating a \emph{smart seat pad}, capturing complex metrics like pedaling stability and leg-strength balance, yielding richer assessments than standard sensors.

Finally, emotional and physiological aspects also strongly shape VR cycling experiences. \citet{potts2024sweating} investigated adaptive VR environments that respond to a user’s emotional state, finding that exertion levels and mood significantly influence how cyclists perceive and interact with virtual contexts.

\subsection{Training Based on Physiological Sensing and HR Zones}

Physiological sensing provides real-time insight into a user’s exertion level and stress, enabling more targeted and effective workouts \cite{seshadri2019wearable}. Yet, as \citet{hao2017mindfulwatch} highlight, few accessible tools offer extensive biosignal data for practical applications like guiding breathing patterns or refining workout strategies.

Most structured training regimens alternate workouts of varying frequency, duration, and intensity. A common approach is to monitor heart rate (HR) within specific \emph{zones}, which systematically relate exertion level to training goals \cite{schneider2018heart}. This allows individuals to build endurance, manage cardiac stress, and reduce injury risk by tailoring their efforts to each prescribed zone. Recent studies have augmented this approach with \emph{music or soundscapes}, using auditory cues to nudge cyclists toward a desired HR. For instance, \citet{rubisch2010} showed that tempo-based music prompts can increase or decrease pedaling cadence, keeping participants closer to target HR zones. Likewise, \citet{fardoun2015} employed ecological soundscapes on stationary bikes, demonstrating that subtle auditory signals can naturally influence cadence and create a more immersive training experience.

Physiologically, each individual’s HR range spans from resting to maximum rates. Between these two endpoints are \emph{zones} corresponding to different exercise intensities and benefits. Various formulas estimate an individual’s maximum heart rate (\( HR_{max} \)), primarily based on age \cite{inbar1994normal, fox1971physical, tanaka2001age, gellish2007longitudinal}; in this work, we use the widely cited equation of \citet{tanaka2001age}, \( HR_{max} = 208 - .7 \times \text{Age} \). Zones are then defined as percentages of \( HR_{max} \). While different zone classifications exist, the five-zone model remains common practice:

\begin{description}
    \item[Zone 1]-- Very light Intensity, 50-60\% of $HR_{max}$.
    \item[Zone 2]-- Light Intensity, 60-70\% of $HR_{max}$.
    \item[Zone 3]-- Moderate Intensity, 70-80\% of $HR_{max}$.
    \item[Zone 4]-- Hard Intensity, 80-90\% of $HR_{max}$.
    \item[Zone 5]-- Maximum Intensity, 90-100\% of $HR_{max}$. 
\end{description}

Because these zones mirror physiological thresholds, ranging from easy endurance work to all-out efforts, aligning exercise intensity to specific zones can optimize cardiorespiratory adaptation while minimizing risk.

\subsection{Summary}

Prior research highlights the potential of physiological sensing data to enhance exergaming by enabling adaptive and personalized workout experiences. However, despite its strong ability to improve exercise quality and engagement, current systems provide little support for amateur users, who often lack actionable feedback to optimize their exertion levels. Biking emerges as a particularly promising use case due to its widespread popularity, seamless integration with physiological sensors, and structured endurance training benefits. Building on these insights, we explore how physiology-driven visualizations can enhance biking exergames by improving user experience and exertion regulation in immersive virtual environments. Our work contributes to the field by (1) leveraging ubiquitously available sensing data for exergaming, (2) addressing the lack of structured physiological feedback for amateur users, and (3) designing adaptive visualizations that support exertion control in VR-based biking. With these foundations, our work explores not just whether bioadaptive exergaming can be effective, but how it might be felt and understood by users in situ.

\section{Study 1: Exploring Suitable HR Zone Visualizations}

Interactive exergames often struggle to guide users toward effective physical activity levels, partly due to limitations in how physiological data (e.g., HR) is presented and interpreted. To address this challenge, we conducted an online survey comparing nine distinct HR zone visualizations, asking participants to rank their preferred designs for future implementation. These visual concepts drew inspiration from mixed reality navigation studies \cite{dey2018systematic,moller2014experimental,mulloni2011handheld}, commercial exergaming platforms \cite{westmattelmann2021exploring,bentvelzen2022tailor}, motivation theory based on visual representations \cite{wagener2023weather}, and Bartle’s four-player types framework \cite{bartle1996hearts}. For each of the nine designs (including a baseline scenario), we produced short (approximately one-minute) demo videos, available in our supplementary material. The methodology of this study followed the approach outlined by \citet{lee2022user}, focusing on user-driven insight into visualization preferences.

\subsection{Method}
We designed an online survey to identify the most preferred visualization.
Inspired by prior work on designing navigation instructions in mixed reality by \citet{lee2022user}, we identified four high-level types of visualizations: Gamification, Change of Environment, Distortion of Reality, and Visual Overlays.
We created nine visualizations, including a baseline scenario, and compared them using a within-subjects design. We showed these visualizations in the form of text and sample video to the participants. Then, they ranked their preferred visualizations per scenario.

\begin{description}
\item[Gamification] aims to engage and motivate individuals by introducing features like points, rewards, competition, and achievement levels (inspired by work of \citet{shaw2016competition,ketcheson2015designing} and \citet{xu2023exploring}).
\item[Change of Environment] involves altering the surroundings or context in which an experience occurs, to influence perceptions, emotions, and behavior (inspired by work of \citet{guo2021benchmarking}).
\item[Distortion of Reality] manipulates perception through visual, auditory, or sensory illusions, allowing users to experience a reality that deviates from actual stimuli(inspired by prior work of \citet{ioannou2019virtual}).
\item[Visual Overlays] blend digital elements with the real environment, enhancing information presentation and providing context (inspired by prior work by \citet{von2022hazard}).
    
\end{description}

Our designs represent and integrate all of these different ways of visualizing information.
For this purpose, we developed two visualizations per possible way of visualizing information, so eight visualizations in total, which are in alphabetical order: \textit{Atmosphere, NPC, Coins, Environment, Frame, GUI, Motion Blur} and \textit{Saturation}. The ninth design is a baseline scenario, where the status quo of cycling with a heart rate sensor connected to your bike computer is portrayed. We present a detailed description of every visualization (cf. \autoref{tab:VisualizationsPics}).


\begin{table*}[t!]
\caption{\emph{Visualizations:} We created eight visualizations for our first study, following four types (gamification, change of environments, distortion of reality, visual overlay), as identified by prior work. For each type, we created two different visualizations. A non-adaptive baseline complements this set. }
\label{tab:VisualizationsPics}
\begin{tabular}{llc|c|c}
\textit{Group}        & \textit{Visualization} & \textit{HR too low}                & \textit{HR optimal}                & \textit{HR too high}               \\ \hline
Change of Environment & Atmosphere             & \includegraphics[width=0.19\textwidth]{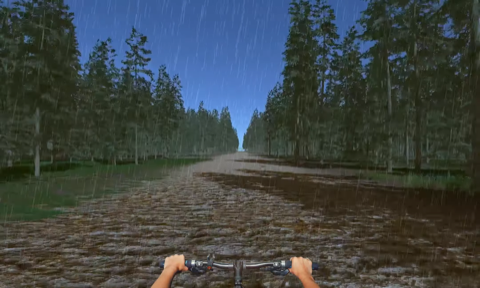}  & \includegraphics[width=0.19\textwidth]{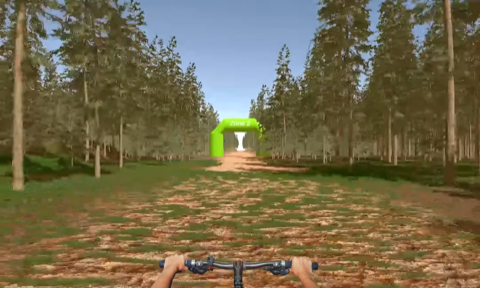}  & \includegraphics[width=0.19\textwidth]{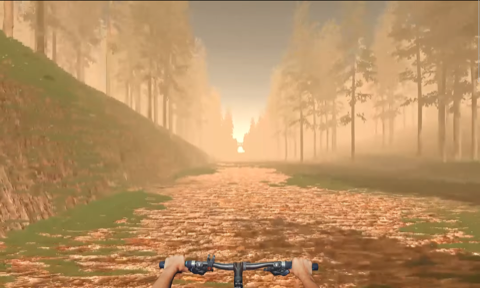}  \\ \cline{2-5} 
                      & Environment            & \includegraphics[width=0.19\textwidth]{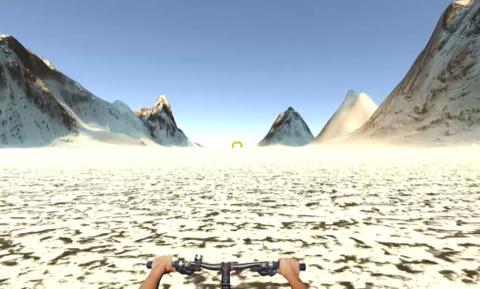}  & \includegraphics[width=0.19\textwidth]{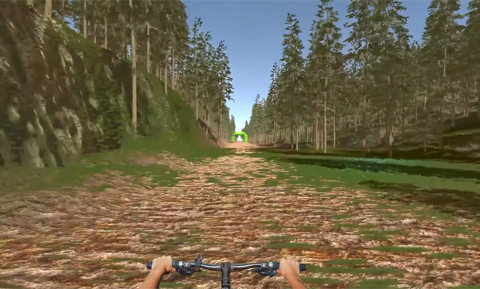}  & \includegraphics[width=0.19\textwidth]{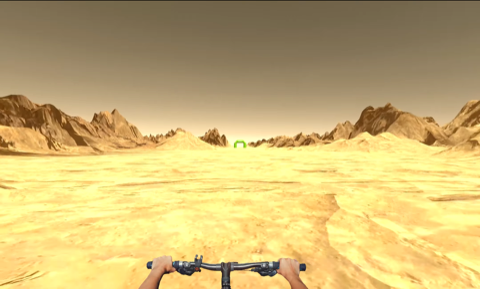}  \\ \hline
Distortion of Reality & Motion Blur            & \includegraphics[width=0.19\textwidth]{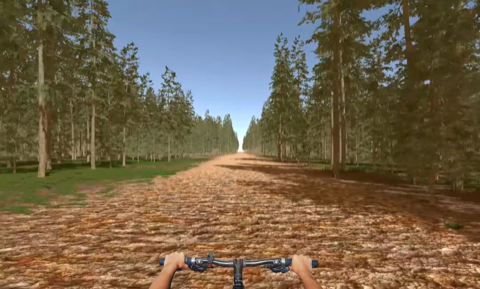}  & \includegraphics[width=0.19\textwidth]{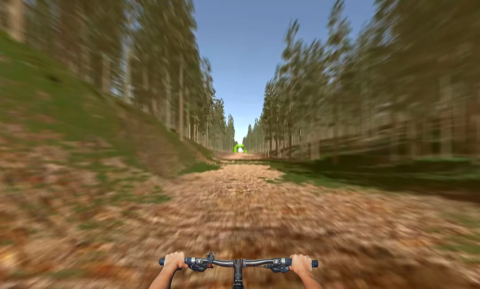}  & \includegraphics[width=0.19\textwidth]{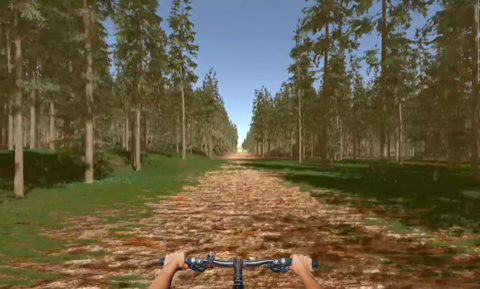}  \\ \cline{2-5} 
                      & Saturation             & \includegraphics[width=0.19\textwidth]{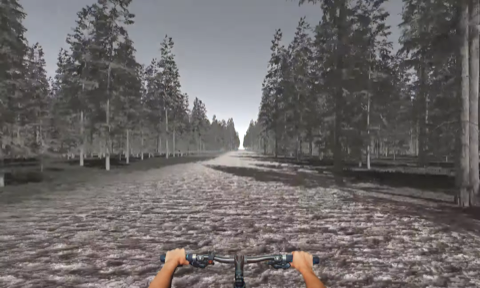}  & \includegraphics[width=0.19\textwidth]{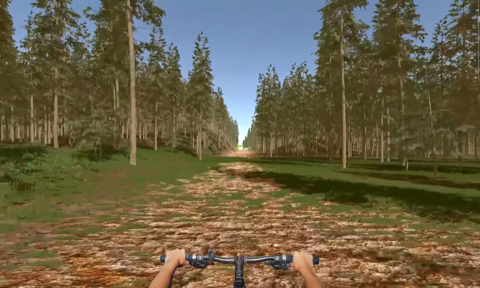}  & \includegraphics[width=0.19\textwidth]{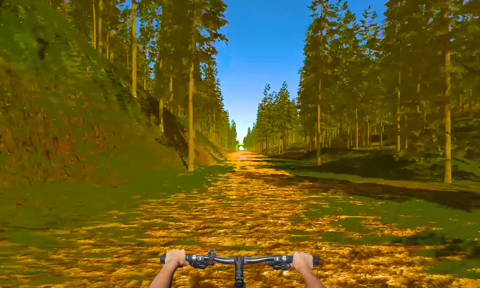}  \\ \hline
Gamification          & NPC                 & \includegraphics[width=0.19\textwidth]{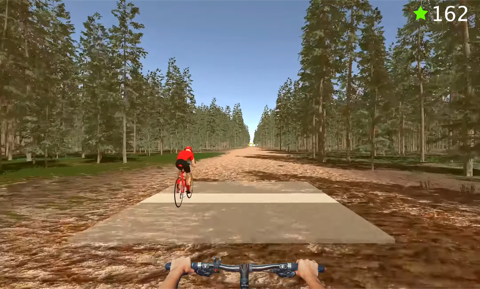}  & \includegraphics[width=0.19\textwidth]{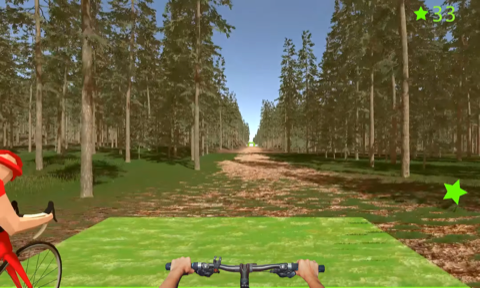}  & \includegraphics[width=0.19\textwidth]{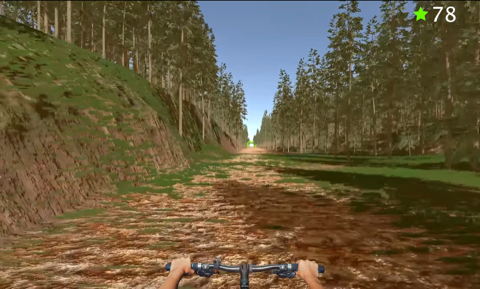}  \\ \cline{2-5} 
                      & Coins                  & \includegraphics[width=0.19\textwidth]{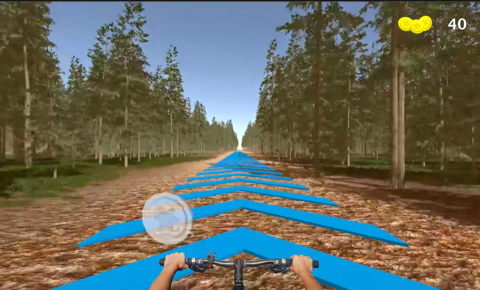}  & \includegraphics[width=0.19\textwidth]{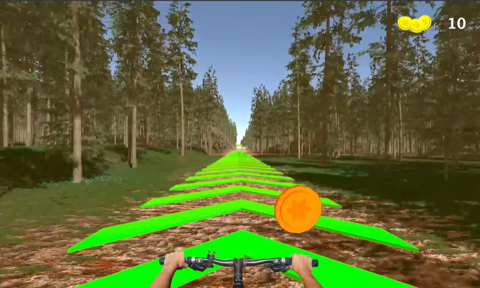}  & \includegraphics[width=0.19\textwidth]{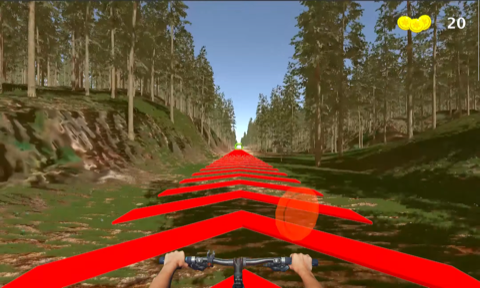}  \\ \hline
Visual Overlays       & Frame                  & \includegraphics[width=0.19\textwidth]{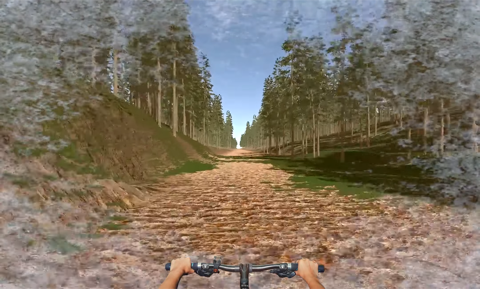}  & \includegraphics[width=0.19\textwidth]{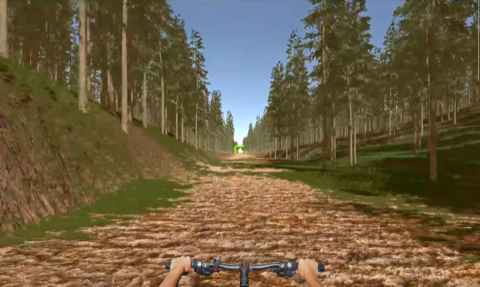}  & \includegraphics[width=0.19\textwidth]{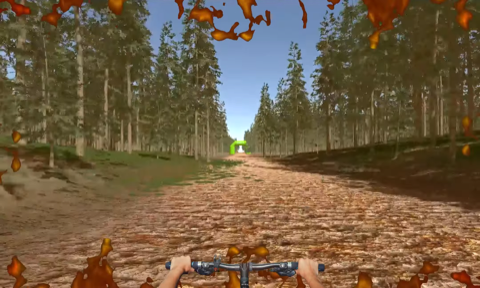}  \\ \cline{2-5} 
                      & GUI                    & \includegraphics[width=0.19\textwidth]{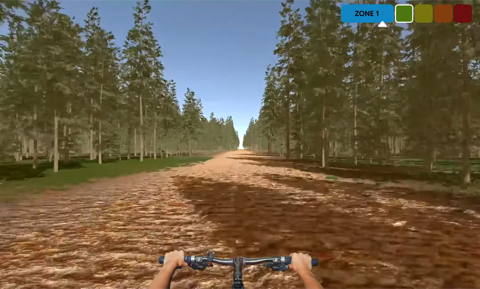}  & \includegraphics[width=0.19\textwidth]{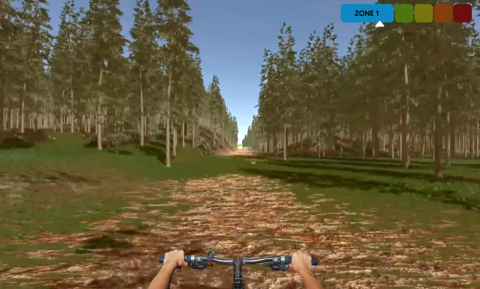}  & \includegraphics[width=0.19\textwidth]{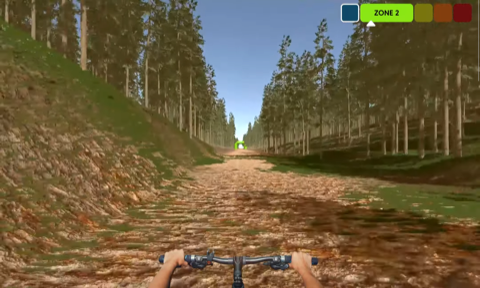} \\ \hline
Baseline &  &    \includegraphics[width=0.19\textwidth]{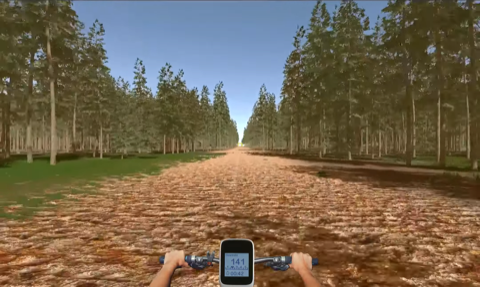}  & \includegraphics[width=0.19\textwidth]{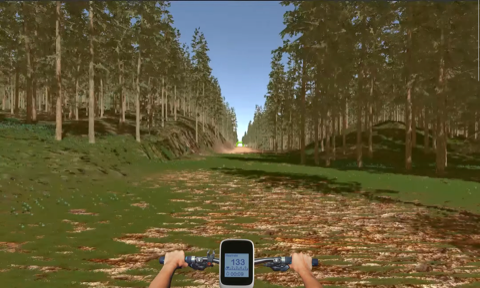}  & \includegraphics[width=0.19\textwidth]{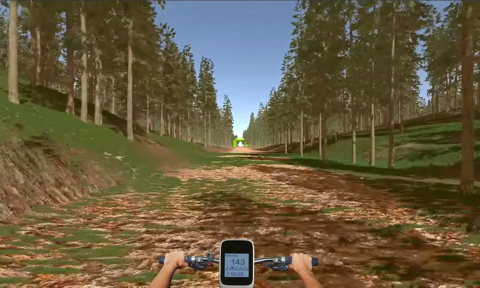}  \\ 
\end{tabular}
\end{table*}

\subsubsection{Change of Environment}
\vspace{-1mm}  \begin{description}
  \item[Atmosphere.]
    In this visualization, atmospheric conditions shift dynamically with the user’s current heart rate. The default (optimal) state is a cool, sunny forest in the afternoon, signifying the target HR zone. If the user’s heart rate exceeds that range, the simulation transitions into an early-morning scene with reddish light and limited visibility due to fog. Conversely, when the user’s heart rate falls below the target, it becomes a rainy, gloomy evening, reinforcing the sense of cold and wind. This design communicates zone deviations through color, lighting, and weather changes, prompting users to adjust their efforts.
    
    \item[Environment.]
    Here, the landscape biome adapts to the user’s heart rate. Under optimal conditions, cyclists remain in a spring forest, visually linked to a moderate, sustainable HR. Surpassing the upper threshold transports riders into a hot, arid desert setting, whereas dropping below the zone places them in an icy, snow-covered environment. By jumping between these contrasting worlds, the system underscores deviation from the target zone, motivating users to adjust pace or intensity to return to the forest.

       \end{description}

\subsubsection{Distortion of Reality}       
    \vspace{-1mm}   \begin{description}\vspace{-1mm}
 \item[Motion Blur.]
    In this design, motion blur intensifies as the user’s heart rate approaches the target zone. The goal is to remain in a “flow mode” where trees and scenery seem to fly by, conveying a sense of high speed. If the user drifts away from the optimal heart rate, the motion blur subsides, making the environment appear slower.

    \item[Saturation.]
    Here, the environment’s color saturation indicates whether the user is in the correct heart rate zone. Under optimal conditions, the surroundings look naturally colored. If the user’s heart rate exceeds the target zone, colors become oversaturated; if it falls below, saturation gradually drains until the world is nearly black and white. These visual changes alert users to adjust their intensity and return to a natural palette.

       \end{description}

\subsubsection{Gamification}     

\begin{description}\vspace{-1mm}
\item[NPC.]
    In this design, an NPC cycles alongside the user on the same route. The NPC’s distance depends on the user’s heart rate relative to a target zone. Ideally, the user remains parallel to the NPC (i.e., they match speed). If the NPC pulls ahead, the user’s heart rate is too low; if it trails behind, the user’s heart rate is too high. A semi-transparent green area and a green line highlight the NPC’s ideal position. Users earn points whenever their front wheel lines up with the green line, signifying alignment with the target HR. If the user drifts away from this zone, the line and area fade to gray, and no points are awarded.

    \item[Coins.]
    Users collect coins or jewels placed along the route, but only if they maintain their heart rate within the specified zone. Whenever the user is outside the target HR, the collectibles turn gray and become uncollectable. To help users gauge their HR status, colored arrows at the bottom of the scene reflect the current zone: green indicates that the user is on target, red signals a heart rate that is too high, and blue signifies that it is too low. These immediate visual cues encourage frequent adjustment of pedaling intensity to maximize coin collection.

       \end{description}
       
\subsubsection{Visual Overlays}          
       \begin{description}\vspace{-1mm}
\item[Frame.]
    In this design, the user aims to keep an unobstructed view by maintaining an optimal heart rate. When the heart rate is within the desired range, the screen remains clear. If the heart rate rises too high, flames appear at the periphery of the user’s field of vision and creep inward with further deviation. Conversely, if the heart rate falls below the target range, an ice-like visual overlay begins to form around the edges, progressively narrowing the user’s central view. These visual cues compel the user to adjust their effort to restore a clear field of view.

    \item[GUI.]
    A color-coded scale is fixed to the right side of the user’s view, depicting multiple heart rate zones from low (left) to high (right). A white arrow beneath this scale indicates the user’s current heart rate. If the arrow aligns with the highlighted zone, the user is in the target range. Deviating above or below causes the wider highlight segment to shift to the corresponding zone, visually illustrating how far the user’s heart rate has strayed. A distinct frame around the optimal segment reinforces the target zone, reminding the user to modify their intensity to keep the arrow on target. We acknowledge that the GUI condition, while informative, was not fully optimized for VR. Future work could investigate more immersive and ergonomically integrated GUI designs.
           
       \end{description}

\subsubsection{Baseline}   
The baseline visualization mimics a typical bike computer mounted on the handlebars. A numerical display presents the user’s current heart rate, while five adjacent boxes (labeled 1 to 5) represent heart rate zones. The box containing a heart icon corresponds to the zone the user is currently in. This design provides straightforward numeric feedback but lacks adaptive visual cues for how to regain or maintain the target zone. The baseline condition intentionally offers minimal feedback to serve as a reference point against which other designs can be evaluated.

\subsection{Apparatus}
We disseminated an online survey to reach a broad participant base via our university mailing list and by directly contacting local cycling clubs. The survey featured brief text descriptions and short, one-minute demo videos showcasing each of the nine different HR zone visualizations. To produce these videos, we first recorded footage from our \emph{Baseline} Unity scene, which depicts a cyclist riding along a straight, forested gravel path (see \autoref{tab:VisualizationsPics}). We overlaid each proposed visualization in Adobe Premiere Pro CC, ensuring a consistent riding scenario across all conditions.

\subsection{Procedure}
We hosted the survey on \textit{Google Forms}\footnote{\url{https://www.google.de/intl/de/forms/about/}, last accessed \today} and publicized it through the university mailing list, a departmental Slack channel, and a local cycling club. After providing informed consent and completing a short demographic questionnaire, participants were presented with each of the nine visualizations in a randomized sequence. They read a concise textual explanation and watched a corresponding video for each visualization before ranking their favorite designs on the basis of personal preference. We did not specify particular usage contexts (e.g., casual exercise or professional training); participants were simply asked to choose which visuals they liked best.

\subsection{Participants}
We received 50 valid survey responses over one week. Participants ranged in age from 20 to 62 ($M_{age}=33.06$, $SD_{age}=10.53$), with 48\% identifying as male (24), 42\% as female (21), and 10\% opting not to specify their gender (5). When asked about previous exposure to AR/VR technologies, most indicated having \textit{never} used these devices (26), followed by \textit{a few times} (13), \textit{once} (7), \textit{weekly} (2), and \textit{daily} (2).  
Regarding cycling habits, participants reported biking an average of 2.22 days per week ($SD=1.86$), covering approximately 37.00 km ($SD=48.97$) weekly. The overall sample thus encompassed a wide range of VR familiarity and cycling experience. 

\subsection{Results}
\begin{figure*}[t]
    \centering
    \includegraphics[width=.6\textwidth]{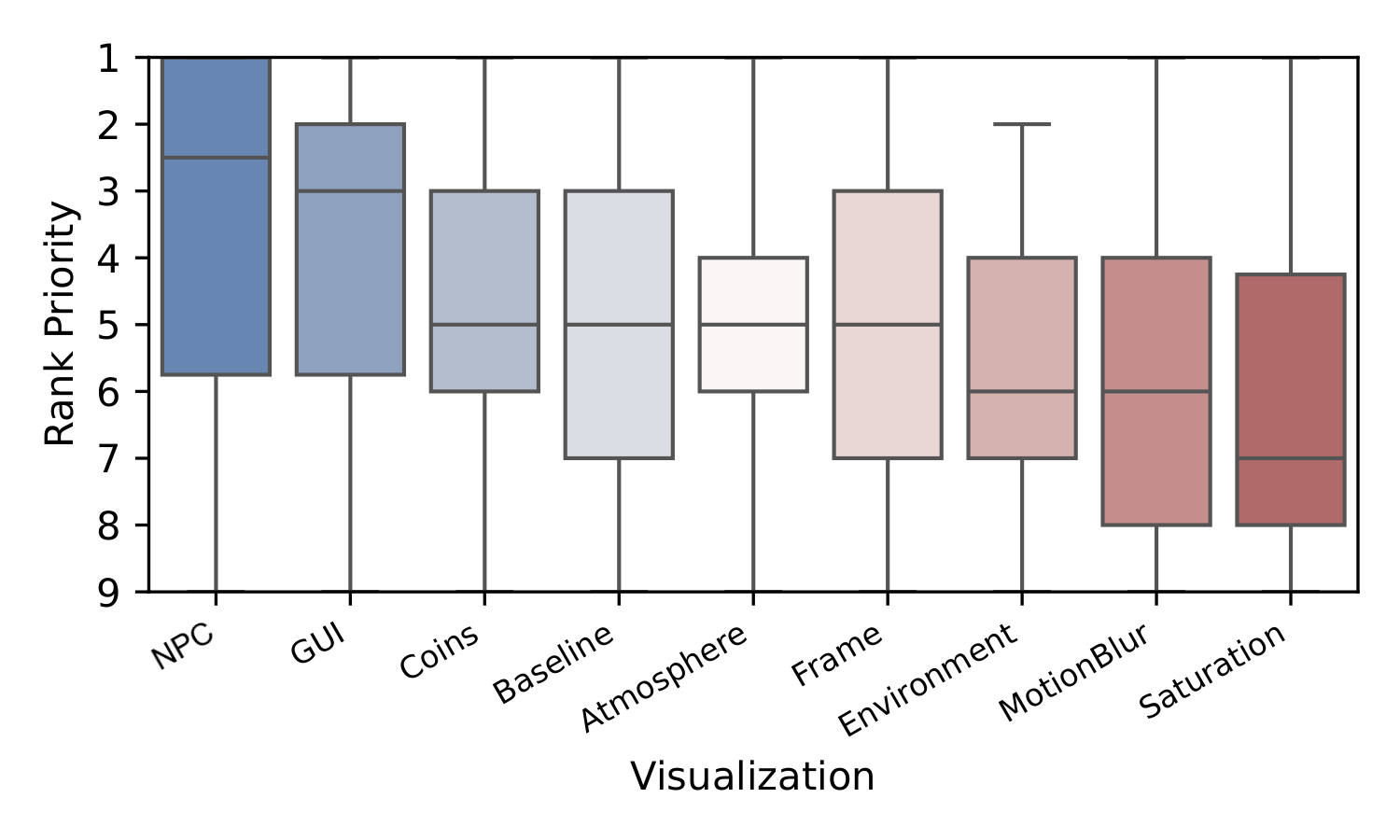}
    \vspace{-6mm}
    \caption{\emph{Rank Preferences Results:} We investigated preference results by computing rank average based on rank sum. Boxplots depict ranks by participants on individual visualizations. Lower scores mean higher preference. Here, the NPC visualization was the most preferred by users across scenarios (lowest rank). }
    \vspace{-1mm}
    \label{fig:study1_results_pref}
\end{figure*}

We began our analysis by computing the average rank for each of the nine visualization concepts, identifying the design with the \emph{lowest} mean rank as the most favored. In this dataset, the top-ranking visualization was ``NPC'' (see \autoref{fig:study1_results_pref}). To verify that these differences in preference were statistically robust, we conducted a Friedman test. This revealed a significant effect of visualization on ranked preference, $\chi^2(8, 50) = 40.23$, $p < .001$, indicating that participants did not view all designs equally. 

\paragraph{Preference Estimation.}
Post-hoc inspection of the mean ranks showed that ``NPC'' (M = 3.60, SD = 2.96) outperformed the other conditions, confirming it as the most popular choice among participants. \autoref{tab:rank_vis} summarizes the descriptive statistics for all nine visualizations.

\begin{table}[h!]
\centering
\caption{Descriptive Statistics for Visualizations}
\small
\label{tab:rank_vis}
\vspace{-4mm}
\begin{tabular}{@{}lcc@{}}
\toprule
Visualization & Mean Rank & SD   \\ \midrule
GUI           & 3.74      & 2.38 \\
Motion Blur   & 5.60      & 2.68 \\
Baseline      & 5.08      & 2.76 \\
NPC        & 3.60      & 2.96 \\
Saturation    & 6.32      & 2.18 \\
Frame         & 5.24      & 2.51 \\
Atmosphere    & 5.18      & 2.17 \\
Environment   & 5.48      & 2.13 \\
Coins         & 4.76      & 2.35 \\ \bottomrule
\end{tabular}
    \vspace{-4mm}
\end{table}

\subsection{Discussion}
Overall, participants rated \textit{NPC}, \textit{GUI}, and \textit{Coins} considerably higher than the other visualizations (\textit{Atmosphere}, \textit{Frame}, \textit{Environment}, \textit{Motion Blur}, and \textit{Saturation}), with \textit{NPC} emerging as the most preferred. We attribute these outcomes to users’ familiarity with straightforward information displays (\textit{GUI}) and gamification elements (\textit{NPC}). This is consistent with prior findings that pre-exposure to conventional interaction paradigms can influence preference in MR \cite{cheng2021semanticadapt,lee2022user}.

Although multiple design directions were initially explored, we aimed to isolate a single, well-received approach for deeper scrutiny. The results suggest \textit{NPC}, and to some extent \textit{GUI}, present a strong candidate for continued development in exergaming scenarios. Nevertheless, designs like \textit{Motion Blur} or \textit{Environment} may excel under different conditions or when combined with other feedback mechanisms, highlighting future opportunities for research. 

We emphasize that each visualization or combination thereof would require a dedicated investigation beyond the current study’s scope. Future work should systematically test these alternatives and consider factors like cognitive load, motion sickness, or real-time adaptation logic to refine how physiological data are conveyed in VR exergames.

\vspace{1mm}\noindent\textbf{Limitations.} This study does not encompass the entire spectrum of possible designs, and the concepts were specifically created for a cycling context, other sports may have distinct demands and motion profiles that limit direct applicability. Future research could expand this design space across different athletic activities to more thoroughly assess how varied visualizations affect both user experience and performance outcomes.
Although combining multiple visual feedback methods might produce richer, more adaptive exergames, it also introduces potential risks such as cognitive overload or technical complexity. Our focus on single-visualization prototypes was intended to preserve clarity and interpretability, allowing us to evaluate each concept’s impact in isolation. We recommend a stepwise approach, validating individual designs thoroughly before layering more complex interactions or visuals.
The evaluation of the nine visualizations was conducted without participants cycling simultaneously, which may have influenced their preferences and led to an early focus on the NPC version.
Despite these constraints, our findings highlight the potential of the tested visualizations, \textit{NPC} in particular, to enhance performance and engagement in exergames. Future work could build on this foundation by refining and integrating designs into other fitness scenarios.

\section{Study 2: Physiologically-Adaptive Visualizations for Biking Engagement}

\subsection{Study Design}

We implemented the highest-rated design from \emph{Study 1}, i.e.,  \emph{Adaptive NPC}, within a 3D VR environment. This study assessed whether a physiologically adaptive NPC, driven by real-time heart rate (HR) monitoring, helps users maintain optimal HR zones while cycling, and how it influences subjective exertion, enjoyment, and motivation.

\paragraph{Conditions.}
To isolate the influence of adaptive feedback, we included three conditions:
\emph{Adaptive NPC}, \emph{Random NPC}, and a \emph{Baseline} control. In the \emph{Adaptive NPC} condition, the system continuously processed each participant’s HR data, identifying which zone they were in and adjusting the NPC’s speed accordingly. By contrast, the \emph{Random NPC} condition presented a gamified element without adaptation, allowing us to distinguish NPC-based gamification effects from physiological adaptation. The \emph{Baseline} condition served as a no-NPC reference.

\paragraph{Design.}
We employed a within-participants experimental setup in which each participant experienced all three conditions in a counterbalanced order. To mitigate learning effects, we used a balanced Latin Williams square design with six distinct orderings \cite{wang2009construction}. In total, the experimental procedure spanned four blocks, as depicted in \autoref{fig:procedure}.
The research protocol was reviewed and approved by our university’s ethics committee, ensuring compliance with ethical standards for human-participant studies.

\vspace{1mm}\noindent We explored the following research questions, informed by related work:
\vspace{0mm}
\begin{description}
      \item [RQ1.] Does real-time adaptive feedback support users in maintaining target HR zones?
      \item [RQ2.] How does adaptive feedback compare to non-adaptive or random designs in performance and experience?
     \item [RQ3.] What effects do adaptive visualizations have on perceived exertion and motivation?
 \end{description}

We evaluated five aspects: 
(I) Heart Rate, (II) Optimal Heart Rate Ratio, (III) intrinsic motivation via the Intrinsic Motivation Inventory (IMI) \cite{ryan2000self}, (IV) subjective exertion via the Perceived Exertion (Borg Rating of Perceived Exertion Scale) \cite{williams2017borg,borg1982psychophysical}, and (V) physical activity enjoyment via the Physical Activity Enjoyment Scale (PACES) \cite{kendzierski1991physical}.
We introduced a \emph{Randomized NPC} condition in which the NPC's speed changed arbitrarily, independent of the participant’s HR to control for any biases in performance and subjective ratings. This setup helps distinguish the effects of genuine physiological adaptation from mere gamification's effects.

Since our main goal was to determine whether participants could more accurately maintain the designated HR zone, we relied on \emph{normalized heart rate} in our analysis. Specifically, each participant’s HR was expressed as a percentage of their individual $\mathrm{HR_{\max}}$, allowing for direct comparisons among users with differing fitness levels.

\subsection{Architecture of the Adaptive Visualization}
\begin{figure*}[t]
    \centering
    \includegraphics[width=0.9\linewidth]{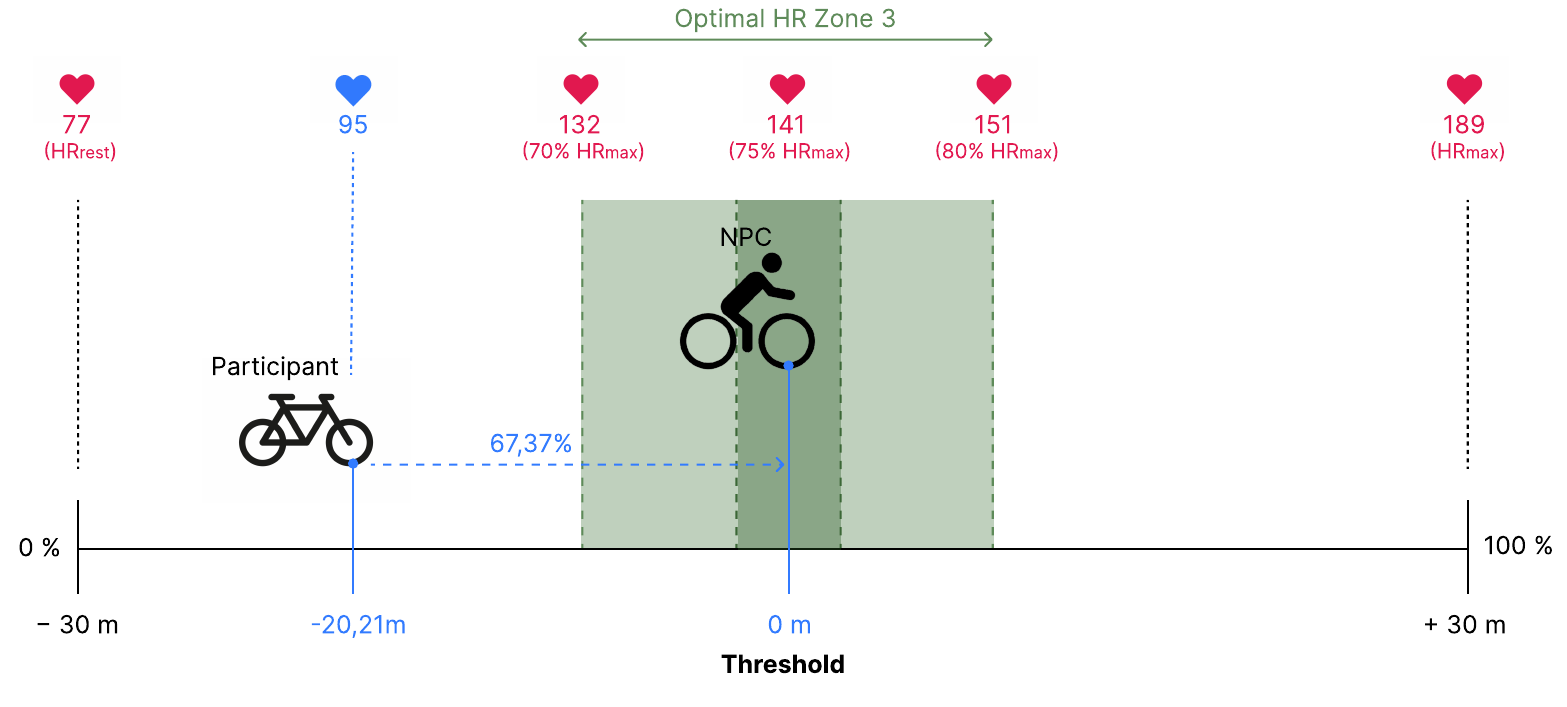}
\vspace{-4mm}
    \caption{\emph{Logic of the Adaptive System.} This diagram illustrates how our VR cycling system uses continuous heart rate (HR) monitoring to adjust an NPC’s speed and position relative to the participant. When the user’s HR is within the target range (e.g., Zone 3: 132--151\,bpm or 70--80\% of HR\textsubscript{max} for a 30-year-old), the NPC maintains pace alongside the user. Should the user’s HR drop below or rise above this range, the NPC lags behind or moves ahead by up to $\pm 30$ meters to reflect the extent of HR deviation. This adaptive mechanism motivates participants to match the NPC’s position, helping them stay within their ideal exercise zone.
}
\vspace{-4mm}
    \label{fig:adaptive_arch}
\end{figure*}

To obtain heart rate data, we first captured raw ECG signals and streamed them to a Python-based TCP/IP server, allowing bidirectional communication between the Lab Streaming Layer (LSL) and our VR (Unity) environment. Real-time ECG preprocessing was handled by the Neurokit Python Toolbox \cite{makowski2021neurokit2} within this client--server pipeline. Specifically, the ECG data passed through a 3--45\,Hz Finite Impulse Response (FIR) band-pass filter (3rd order) before Hamilton’s method \cite{hamilton2002open} segmented the signal to detect QRS complexes. The system then calculated mean heart rate (HR) from these detected peaks, providing the necessary real-time physiological data for our adaptive exergaming logic.
Our ECG pipeline did not include explicit motion artifact filtering. We note this as a technical limitation and that future systems could be improved.

\subsection{Independent Variables}

We implemented three conditions to disentangle the effects of physiological adaptation from those of gamification. In addition to a \emph{Baseline} (no NPC) and an \emph{Adaptive NPC}, we added a \emph{Random NPC} whose speed changed unpredictably. This design lets us gauge whether improvements in performance measures (e.g., maintaining target heart rate, motivation) stem from adaptive feedback or simply from riding alongside any NPC. Comparing the \emph{Random NPC} to the \emph{Adaptive NPC} isolates the specific impact of real-time physiological adaptation.

Participants in all three conditions followed the same procedure: they were immersed in a high-fidelity VR forest with realistic environmental audio and rode along a straight gravel road. Steering was unnecessary because the path extended forward without turns. To raise their heart rate (HR), participants pedaled more intensely; to lower it, they eased off or stopped pedaling. Over a six-minute session, they aimed to keep their HR within progressively more demanding zones. For the first two minutes, they remained in \emph{Zone~1} (very light intensity, 50--60\% of $HR_{\max}$). After two minutes, they shifted to \emph{Zone~2} (light intensity, 60--70\% of $HR_{\max}$). Following another two minutes, the target increased to \emph{Zone~3} (moderate intensity, 70--80\% of $HR_{\max}$). At the end of the six minutes, a text prompt appeared in their field of view, signaling the completion of the run.

\begin{figure*}[t]
    \centering
    \includegraphics[width=\linewidth]{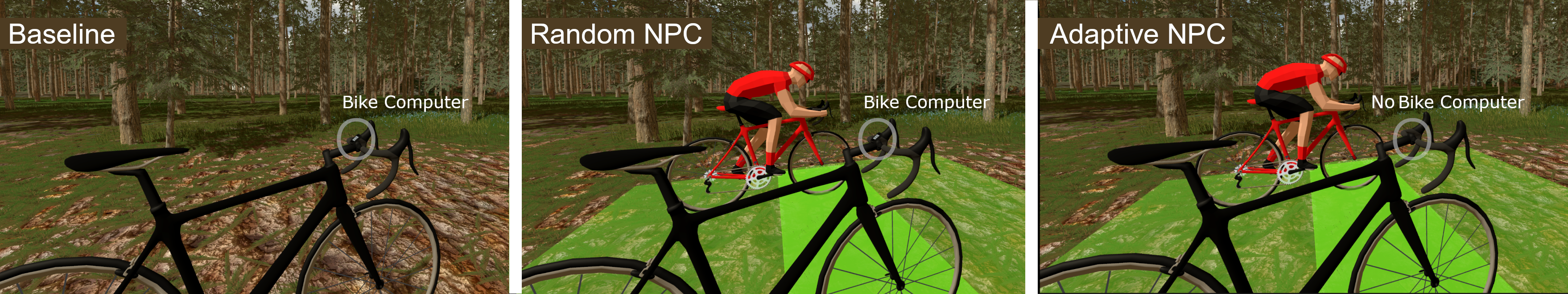}
    \vspace{-6mm}
    \caption{\emph{In-Game Overview of the Three Conditions `Baseline', `Random NPC' and `Adaptive NPC':} In the Baseline condition (left), the participant views a bike computer displaying their heart rate but does not interact with an NPC. In the Random NPC condition (middle), an NPC cycles alongside the participant, but its position changes randomly and is not influenced by the participant’s heart rate. In the Adaptive NPC condition (right), the NPC’s position dynamically adjusts based on the participant’s heart rate, encouraging the participant to maintain their heart rate within the target zone. }
    \vspace{-4mm}
    \label{fig:overviewScenarios}
\end{figure*}

\subsubsection{Baseline}
In the \textit{Baseline} condition, participants cycle along a forested gravel road while aiming to remain within the designated heart rate (HR) zone. A conventional bike computer is mounted on the handlebar, displaying the participant’s current HR and a bar labeled 1--5 (representing the five HR zones). A black heart icon appears over the number corresponding to the user’s current zone, and a green box highlights the \emph{target} zone. The user is in the correct zone if the green box encircles the heart icon. Participants can increase or decrease their HR by pedaling more or less vigorously (or even stopping), to keep the heart icon aligned with the target zone.

\subsubsection{Random NPC}
In this condition, an NPC rides alongside the participant on the same straight forest road. As in the baseline, users aim to maintain a specified HR zone and receive continuous HR feedback from the handlebar bike computer (showing both their current and target zones). However, the NPC’s distance from the participant fluctuates \emph{randomly} within a $\pm 30$\,meter range. Theoretically, these unpredictable movements should not influence the participant’s performance or behavior, allowing us to test whether NPC-based gamification alone affects HR maintenance.

\subsubsection{Adaptive NPC}
Next to the users, an NPC rides the same route on his bike. They aim to keep up with the NPC. This means the NPC should always ride the same height as the user. The user should not overtake the NPC, ride ahead or behind. The distance of the virtual NPC changes depending on the user’s current HR. The NPC's speed is based on the user’s HR and the frequently set HR zone. The users can assume their own HR by the distance to the NPC. If the NPC is in front of the user, the user’s heart rate is too low. If the NPC is behind the user, the user’s heart rate is too high. A light green (transparent) area and a green line are displayed around the NPC (\autoref{fig:adaptive_arch}). If the distance is too great, this area will be grayed out.
In this scenario, there is no bike computer attached to the bike's handlebar. Therefore, the users cannot see their exact current HR and zone. \autoref{fig:overviewScenarios} visually compares the conditions.

\begin{figure*}[t]
    \centering
    \includegraphics[width=.95\textwidth]{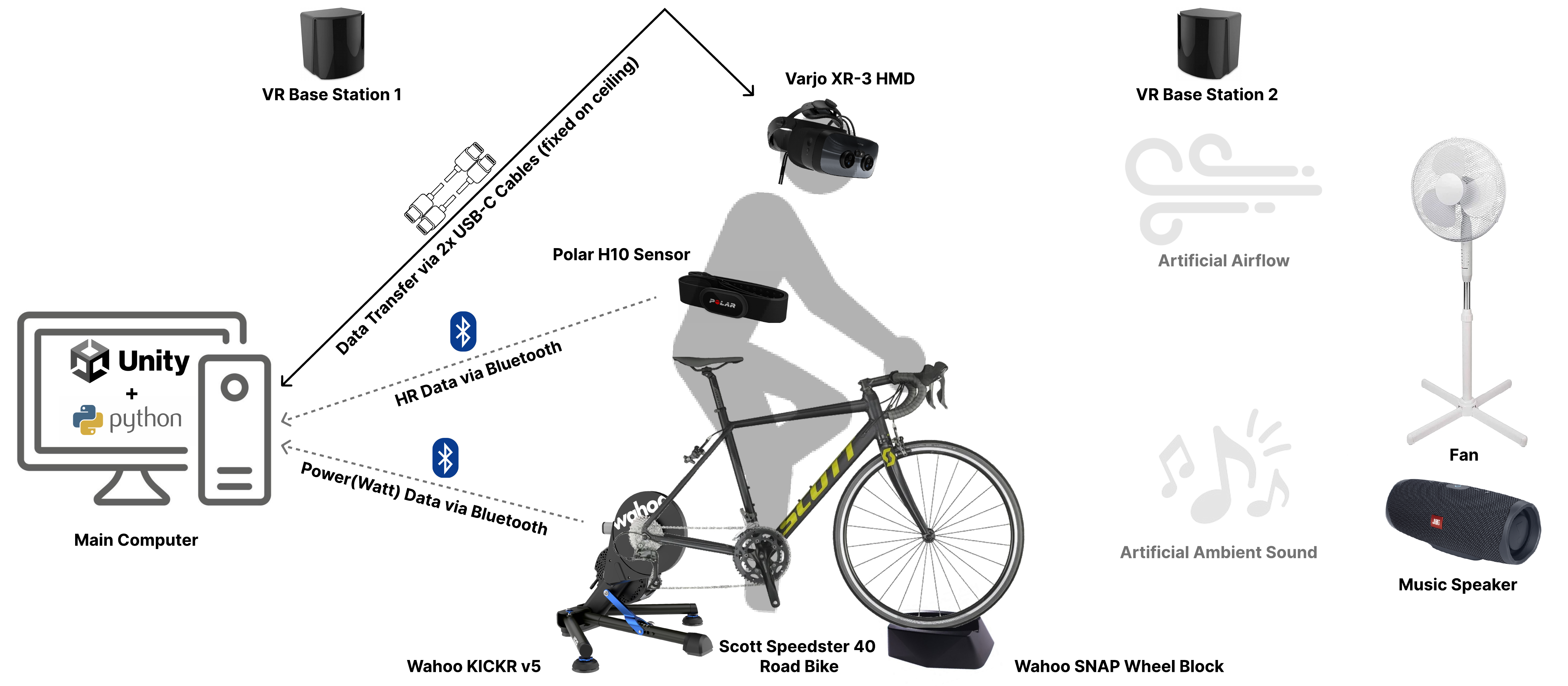}
    \vspace{-4mm}
    \caption{\emph{Setup of our VR Cycling Simulator:} The image illustrates the setup of the VR cycling simulator, which includes a Scott Speedster 40 road bike mounted on a Wahoo KICKR V5 trainer. A Polar H10 sensor monitors heart rate, and the system uses a Varjo XR-3 HMD for VR environment display. Data is transmitted between the bike and the main computer, which runs Unity and Python via Bluetooth. Additional elements include two VR base stations for tracking, a fan for artificial airflow, ambient sound from a music speaker, and a Wahoo SNAP Wheel Block for bike stability. For a complete description of the apparatus, refer to \autoref{sec:apparatus}.
}\vspace{-4mm}
    \label{fig:study2_setup}
\end{figure*}

\subsection{Dependent Variables}

\subsubsection{Heart Rate and Optimal HR Ratio}
HR is a primary physiological indicator of exercise intensity. As detailed in \autoref{sec:hr}, our system continuously recorded participants’ HR to capture their real-time physical responses. We evaluated normalized heart rates to assess how well participants maintained their target HR zone across conditions. As the goal of our system is not high exertion but controlled exertion, we focus on the stability of HR adaptation rather than peak values.
To quantify how well they adhered to their prescribed HR range, we calculated participants' \emph{Optimal HR Ratio}. Specifically, we determined, at each time point, whether the participant’s HR was within the target zone, then converted the proportion of “on-target” time into a percentage. A higher Optimal HR Ratio reflects stronger compliance with the desired intensity level and can be interpreted as more efficient engagement in the prescribed workout.

\subsubsection{Borg Rating of Perceived Exertion (RPE)}
Perceived Exertion (RPE) measures how strenuous participants \emph{feel} their workout is, as opposed to relying purely on physiological readings. We employed the Borg Rating of Perceived Exertion Scale \cite{borg1982psychophysical}, a 6--20 range where 6 corresponds to “no exertion at all” and 20 signifies “maximal exertion.” After each cycling session, participants reported their RPE score. Higher values indicate a greater sense of difficulty, offering insight into subjective workload beyond HR-based metrics.

\subsubsection{Physical Activity Enjoyment Scale (PACES)}
The Physical Activity Enjoyment Scale (PACES) \cite{kendzierski1991physical} is widely used to measure subjective enjoyment of exercise \cite{mullen2011measuring,jekauc2020measurement}. In its original form, PACES contains 18 statements rated on a 7-point Likert scale, capturing the pleasure and satisfaction derived from physical activity. Because 11 of these items are negatively worded, their scores must be reversed before calculating the overall enjoyment level. Higher PACES totals indicate a greater sense of enjoyment during physical exercise.

\subsubsection{Intrinsic Motivation Inventory (IMI)}
The Intrinsic Motivation Inventory (IMI) \cite{ryan2000self} gauges the degree to which participants find an activity inherently rewarding rather than driven by external pressures. This study employed a 30-item version covering five dimensions: \textit{Interest/Enjoyment}, \textit{Perceived Competence}, \textit{Effort/Importance}, \textit{Pressure/Tension}, and \textit{Value/Usefulness}. Some items were negatively worded and thus reverse-scored before summation. Higher scores on each dimension reflect stronger intrinsic motivation within that particular category.

\subsection{Apparatus and Implementation}
\label{sec:apparatus}
We developed our VR cycling environment and all related tasks in Unity 3D (Version~2022.3.20f1), the same platform used to create sample videos in Study~1. For the physical setup, we combined a {Scott Speedster 40} road bike (size~M, 54\,cm) with a {Wahoo KICKR Smart Trainer~v5}. The Scott Speedster features an aluminum frame and fork, plus a Shimano Claris 2$\times$8 gearbox; size~M was chosen to accommodate the average European adult \cite{robinson2015population}.

The Wahoo KICKR v5 was selected for its lateral movement support and auto-calibration. According to the manufacturer, it achieves $\pm 1\%$ accuracy\footnote{\url{https://eu.wahoofitness.com/devices/indoor-cycling/bike-trainers/kickr-buy}, last accessed \today.}. We connected the KICKR to our main computer via Bluetooth (\emph{Cycling Power-Service})\footnote{\url{https://www.bluetooth.com/specifications/specs/cycling-power-service-1-1/}, last accessed \today.}, enabling a Python script to receive live cycling power data (in Watts). Using a standard velocity-conversion method \cite{savadi2022development}, we estimated the participant’s speed in km/h but did not store raw power data. We removed the bike’s rear wheel to mount it directly on the trainer, then secured its front wheel in a Wahoo SNAP Wheel Block (fixed to the floor with duct tape), preventing any steering input in the VR simulation.

Participants wore a \emph{Mixed Reality} head-mounted display (HMD) to experience the immersive environment, while the Unity scene featured a straight gravel road through a forest. Because steering was disabled, forward motion was the sole user input. An \emph{NPC model} (purchased from the Unity Asset Store\footnote{\url{https://assetstore.unity.com/packages/3d/vehicles/land/low-poly-cyclist-184962}, last accessed \today.}) provided a virtual cycling companion where needed (e.g., in \emph{Adaptive NPC} or \emph{Random NPC} conditions). This apparatus and software stack ensured consistent, controlled VR biking scenarios across all conditions.

For head-mounted display (HMD) hardware, we selected a {Varjo~XR-3}, which offers a 115° horizontal field of view, a 90\,Hz refresh rate, and dual 12\,MP video pass-through at 90\,Hz. Its built-in motorized interpupillary distance (IPD) adjustment (59--71\,mm), three-point headband, and face cushions help ensure a secure fit. Although the device weighs approximately 980\,g, we chose an XR-capable HMD instead of a VR-only model for flexible adaptation and the option to quickly remove participants from the virtual environment if they experienced motion sickness. Two {HTC Steam\,VR Base Station~2.0} units enabled outside-in tracking, and we ran two USB-C cables overhead to minimize cable interference.

Participants wore a {Polar~H10} chest strap (130\,Hz sampling rate, Polar, Finland) to capture electrocardiogram (ECG) data. The strap was positioned around the lower chest; after roughly two minutes of adjustment, it produced consistent signal quality. To enhance immersion and reduce motion sickness, we placed a fan in front of the bike to simulate airflow and provided natural forest sound effects (birdsong, rustling leaves) via external speakers \cite{d2017efficacy,shaw2017evaluating}. This minimized the need for additional head-mounted audio equipment and maintained a low-noise environment.

\autoref{fig:study2_setup} illustrates the overall VR cycling simulator. We also implemented an auxiliary GUI window, visible only to the study examiner, to input participant ID and age for accurate HR-zone calculations before each session. While participants pedaled, the examiner could monitor the participant’s current HR, detected HR zone, and remaining time, ensuring real-time supervision and quick adjustments if necessary.

To support replication and further research, we have released our Unity project, including the NPC adaptation logic and heart rate integration scripts, on Open Science Framework (see Section \ref{osf}). This repository includes setup instructions and guidelines for adapting the visualizations. Our pipeline integrates Unity, Python (NeuroKit), and Lab Streaming Layer (LSL), and is designed for modular reuse in varied mixed reality contexts. We welcome extensions and remixing of this toolkit for adaptive applications.

\subsection{ECG Recording and Preprocessing}
\label{sec:hr}
We acquired ECG data at a 130\,Hz sampling rate using a Polar H10 chest strap (Polar, Finland). Before recording, each electrode was moistened with lukewarm water and placed just below the chest muscles, over the xiphoid process of the sternum, ensuring proper contact and minimal noise. All real-time ECG processing occurred via the Neurokit Python Toolbox~\cite{makowski2021neurokit2}. We first applied a 3rd-order Finite Impulse Response (FIR) band-pass filter (3--45\,Hz) to reduce baseline drift and high-frequency artifacts. Hamilton’s method~\cite{hamilton2002open} then segmented the filtered signal to detect QRS complexes, from which the instantaneous heart rate (HR) was extracted.

\subsection{Task}
Building on prior work \cite{kosch2022notibike}, we created a 6-minute cycling task set in a high-fidelity virtual forest. During a 2-minute familiarization phase, participants experienced the environment and learned to operate the road bike. Once the main session began, they pedaled along a straight gravel path for 6 minutes without steering inputs. The objective throughout was to maintain a \emph{target HR} that evolved every 2 minutes: from Zone~1 (very light, 50--60\% of $HR_{\max}$) to Zone~2 (light, 60--70\% of $HR_{\max}$), and finally Zone~3 (moderate, 70--80\% of $HR_{\max}$). Participants received instructions on viewing HR data (and, if relevant, NPC behavior) in their headsets. They could increase HR by pedaling more vigorously or lowering it by reducing effort or briefly stopping. At the 6-minute mark, a text prompt informed them to end the run.

\subsection{Procedure}
\label{sec:procedure}
We conducted the study in a controlled lab environment at our university. Upon arrival, each participant was briefed on the study’s objectives and the HR zones they would aim to maintain, as depicted on a printed reference sheet. We obtained informed consent, emphasizing that participants could halt the study at any point should they experience discomfort or motion sickness. Next, we adjusted the bike saddle to a suitable height, lightly moistened the Polar H10 chest strap, and demonstrated how to position it properly using a provided illustration. For participants unfamiliar with a road bike’s gear shifting, we offered a brief tutorial before they mounted the bike to begin a training phase in VR.

Participants acclimated to the virtual environment during this training session and practiced shifting gears without directly viewing their limbs in VR. Immediately following the training, participants rested briefly and completed a demographics questionnaire, which included items about any motion sickness encountered. They were given a unique ID to preserve anonymity when correlating their survey responses with ECG data. 

The main experiment consisted of three distinct cycling conditions presented consecutively. Before each condition, the experimenter entered the participant’s ID, age, and desired zone duration into a private UI panel to ensure correct HR-zone adaptation. As participants cycled, we matched a fan’s speed to their virtual pace to boost immersion. A short, standardized explanation was read aloud before the start of each condition. Upon finishing a condition, participants completed a corresponding section of the questionnaire.

\begin{figure*}[t!]
    \centering
    \includegraphics[width=\linewidth]{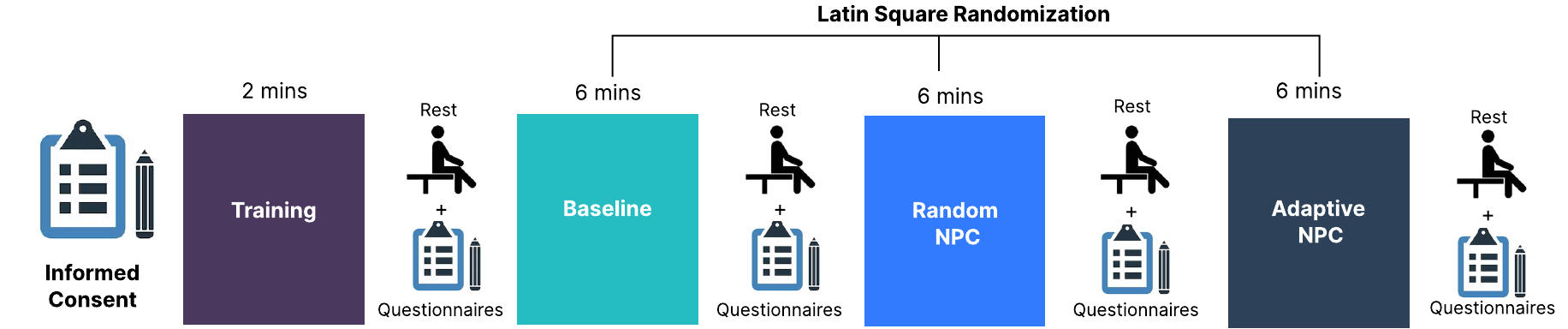}
    \vspace{-5mm}
    \caption{\emph{Experiment Procedure.} Participants begin with a 2-minute training session to get accustomed to the VR environment and cycling equipment. The experiment then proceeds through three 6-minute conditions (Baseline, Random NPC, and Adaptive NPC) in a Latin-square randomized order. After each condition, participants rest briefly and complete questionnaires regarding their experience. The study ends once all three conditions are finished. A more detailed description is provided in \autoref{sec:procedure}.}
    \vspace{-4mm}
    \label{fig:procedure}
\end{figure*}

After experiencing all three conditions, participants were debriefed and compensated for their time. \autoref{fig:procedure} provides an overview of the entire workflow.

\subsection{Participants}
We recruited 18 participants ($M_{\mathrm{age}}=35.27$, $SD_{\mathrm{age}}=12.88$), of whom 12 identified as male (66.67\%) and 6 as female (33.33\%). Initially, we sought individuals who cycle more than 5\,km per week \cite{reddy2010biketastic} or engage in other sports involving HR-based training. However, we also included participants who reported cycling less frequently, yet still on a regular basis.

On average, participants reported cycling $3.06 \pm 2.15$ days per week and covering $77.72 \pm 98.79$\,km weekly. They also engaged in other sports $2.61 \pm 1.38$ times per week. Five participants specifically mentioned using heart rate data to guide their workouts. Regarding AR/VR experience, nine had never used a head-mounted display (HMD), two had done so once, six had tried one a few times, and 1 participant used an HMD weekly. Only 1 of the 18 owned a VR device.

During a 10-point Likert scale assessment of motion sickness in the training phase, 16 participants reported no or minimal discomfort (scores of 1--3), and 2 reported minor to medium symptoms (scores of 4--6). No one gave higher ratings (7--10) or exited the study early, indicating that the VR setup was broadly tolerable across varied experience levels.

\begin{figure*}[t]
    \centering
    \includegraphics[width=\linewidth]{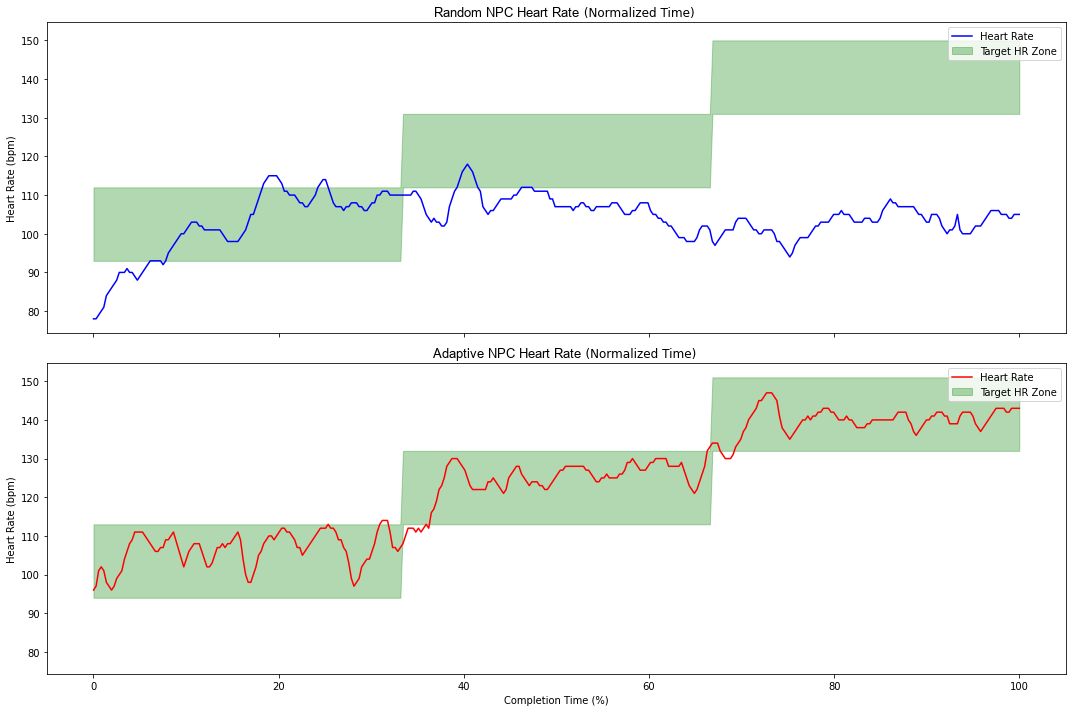}
\vspace{-6mm}
    \caption{\emph{Optimal Heart Rate Adaptation:} We depict the participant's HR evolution across Target HR zones for the two adaptive visualizations. The \textsc{Random NPC} is depicted on top, while the \textsc{Adaptive NPC} is at the bottom. For the \textsc{Adaptive NPC} visualization, participants kept their HR in the optimal HR ratio for $.749$ \% (SD = .02) of the time, while in the \textsc{Random NPC} visualization, participants stayed, on average, $.655$ \% (SD = .021) of their time in the optimal HR ratio. HR data points are fitted within a uniform time scale.}
\vspace{-4mm}
    \label{fig:optimalHRAdaptation}
\end{figure*}

\section{Study 2: Results}
We begin by reporting outcomes for heart rate (HR) and the \emph{Optimal HR Ratio}, followed by the subjective questionnaire measures (BORG, PACES, and IMI). We employed Linear Mixed Models\footnote{Restricted Maximum Likelihood (REML) estimation with Satterthwaite’s approximation for degrees of freedom.} to account for between-participant variance and repeated measures across different target zones. Specifically, for ECG-derived measures, we used:
\begin{center}
\texttt{measure ~ Condition + (1 | participant) + (1 | Target Zone)}
\end{center}
where \texttt{Condition} (Adaptive NPC, Random NPC, Baseline) was a fixed effect, and both \texttt{participant} and \texttt{Target Zone} were random intercepts. For the subjective questionnaires (BORG, PACES, and IMI), our model simplified to:
\begin{center}
\texttt{measure ~ Condition + (1 | participant)}
\end{center}
since target-zone considerations did not apply to these self-report data. We report standardized beta coefficients (\texttt{Std. $\beta$}) to describe effect sizes independently of each variable's original scale \cite{wiley2019statistical, brysbaert2018power}, providing a clear sense of the relative magnitude of each effect.

\begin{figure*}[t]
  \centering
  \newcommand{\panelheight}{0.24\textheight} 

  \begin{subfigure}[t]{0.49\textwidth}
    \centering
    \includegraphics[width=\linewidth,height=\panelheight]{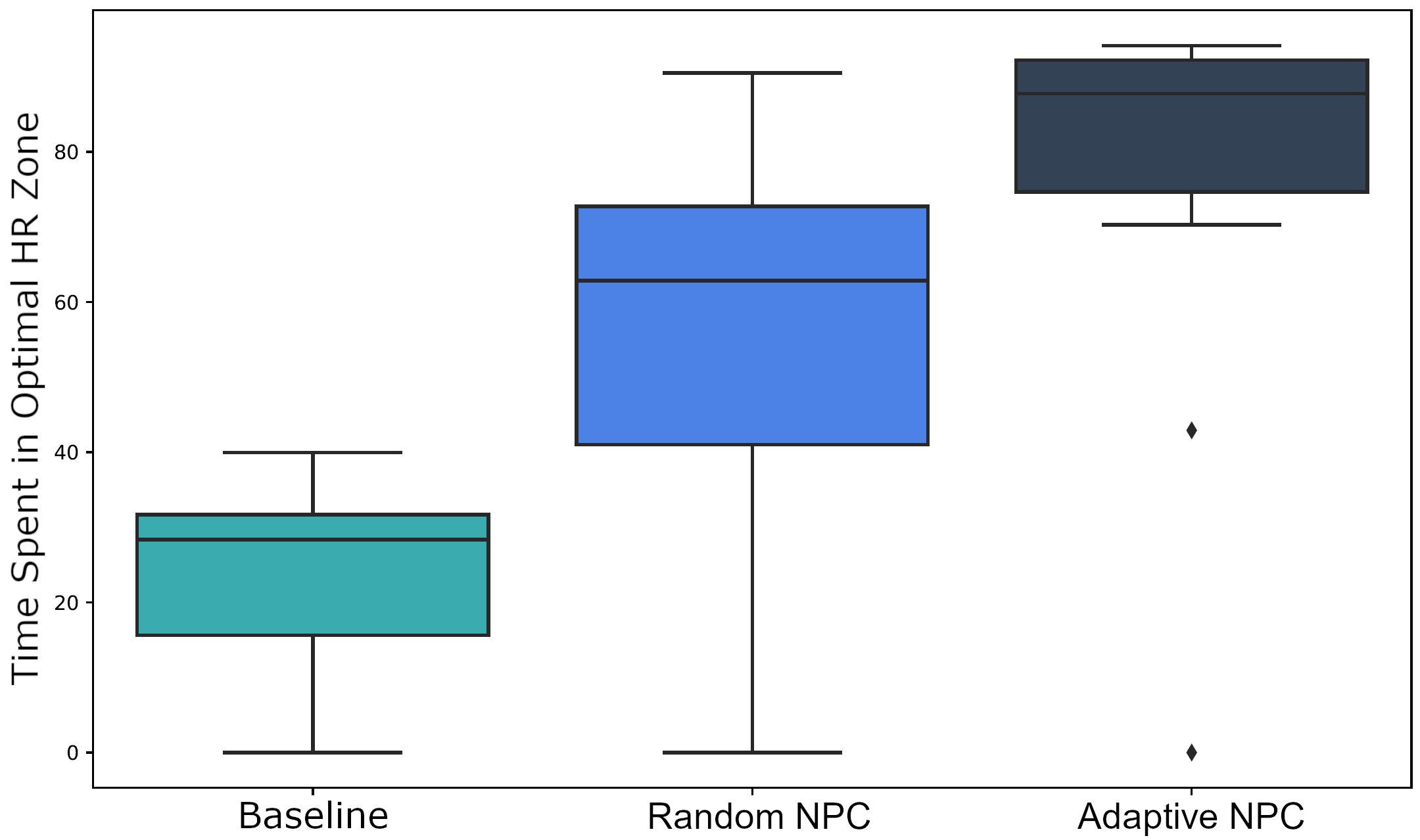}
    \caption{Time in optimal HR zone.}
    \label{fig:results_hr_ratio}
  \end{subfigure}\hfill
  \begin{subfigure}[t]{0.49\textwidth}
    \centering
    \includegraphics[width=\linewidth,height=\panelheight]{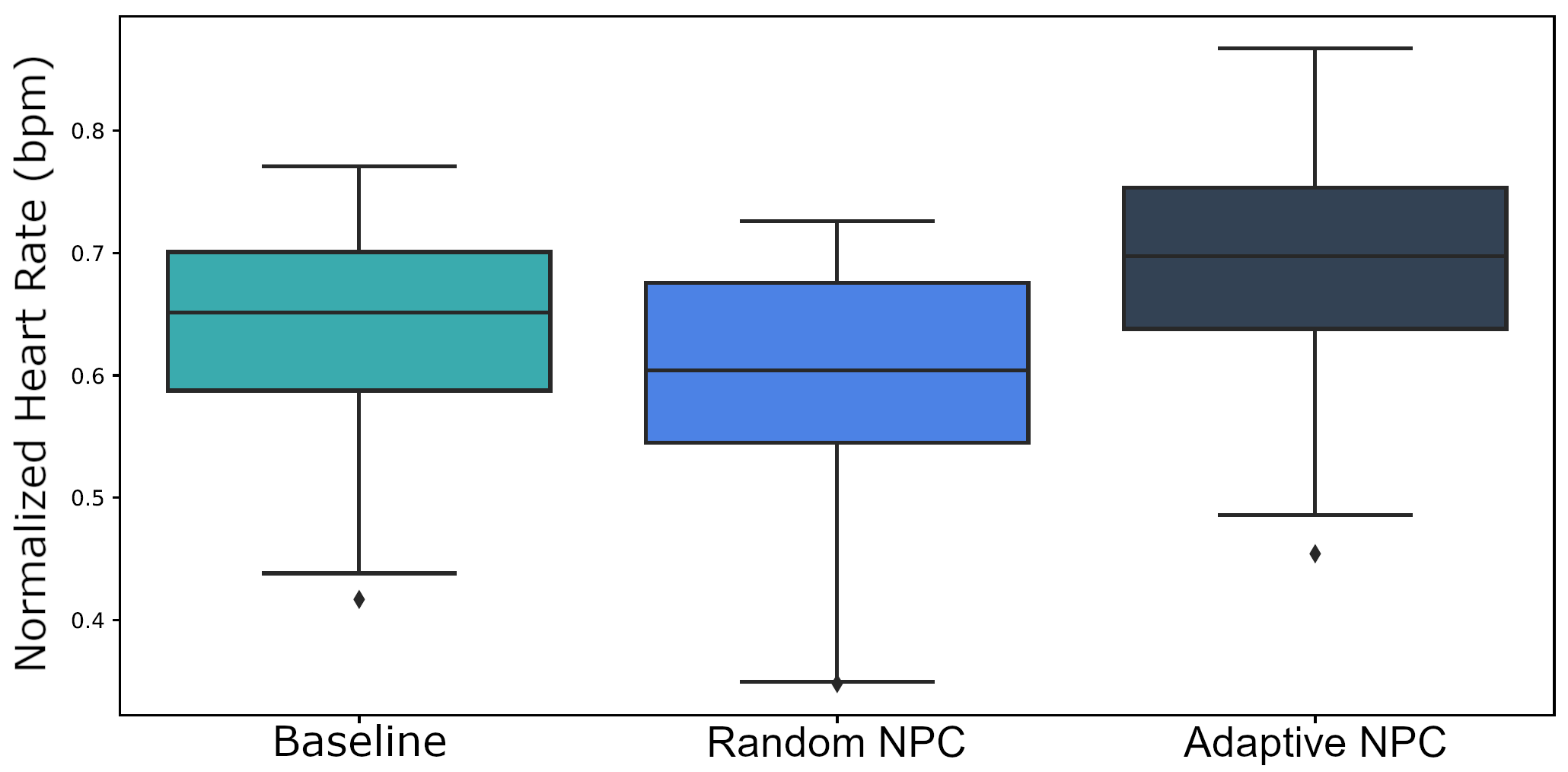}
    \caption{Normalized HR.}
    \label{fig:results_hr}
  \end{subfigure}

 \caption{Comparison of heart-rate outcomes across conditions. 
(a) Time in optimal HR zone. Participants in the \textsc{Adaptive NPC} condition spent significantly more time in the optimal HR zone compared to both \textsc{Baseline} and \textsc{Random NPC}. While both control conditions reduced time in the target zone, the reduction was strongest in \textsc{Baseline}, indicating that the adaptive feedback provided the most consistent support for sustaining optimal exertion. 
(b) Normalized HR. The \textsc{Adaptive NPC} condition maintained a significantly higher heart rate than both \textsc{Baseline} and \textsc{Random NPC}. The \textsc{Random NPC} condition produced the lowest normalized HR overall, reflecting a substantial drop in participants’ engagement compared to the adaptive support. 
Together, these results show that adaptive feedback was most effective in promoting and maintaining active cardiovascular engagement.}
  \label{fig:combined_hr}
\end{figure*}

\subsection{ECG}
\subsubsection{Heart Rate}

Our heart rate model exhibited a moderate overall explanatory power ($R^2_c = .26$), with the fixed effects contributing $R^2_m = .14$. The intercept, corresponding to the \textsc{Adaptive NPC} condition, was estimated at 0.69 (95\% CI [.65, .73], $t(138) = 32.37$, $p < .001$). Within this framework, the effect of the \textsc{Baseline} condition was significant and negative ($\beta = -.05$, 95\% CI [-.08, -.01], $t(138) = -2.80$, $p = .006$), with a standardized effect size of $-.50$ (95\% CI [-.86, -.15]), indicating a medium effect \cite{brysbaert2018power}.

Likewise, the effect of the \textsc{Random NPC} condition was significant and negative ($\beta = -.09$, 95\% CI [-.12, -.06], $t(138) = -5.20$, $p < .001$), with a larger standardized effect size of $-.94$ (95\% CI [-1.29, -.58]), suggesting a large effect \cite{brysbaert2018power}. These findings imply that \textsc{Baseline} participants had a modestly lower normalized HR than those in \textsc{Adaptive NPC}, while \textsc{Random NPC} yielded the \emph{lowest} normalized HR overall (see \autoref{fig:results_hr}). The relatively large negative effect size in the \textsc{Random NPC} condition underscores its stronger influence on reducing participants’ HR, as evidenced by the more pronounced departure from the intercept estimate.

\subsubsection{Optimal Heart Rate Ratio}

The model’s intercept was estimated at 79.08 (95\% CI [68.19, 89.97], $t(46) = 14.62$, $p < .001$). Within this model, the \textsc{Baseline} condition exerted a significantly negative influence on the percentage of time spent in the optimal HR zone, with a large effect size ($\beta = -56.02$, 95\% CI [-68.49, -43.54], $t(46) = -9.04$, $p < .001$; Std.~$\beta = -1.76$, 95\% CI [-2.15, -1.36]). Likewise, the \textsc{Random NPC} condition also produced a significantly negative effect ($\beta = -22.57$, 95\% CI [-35.04, -10.09], $t(46) = -3.64$, $p < .001$; Std.~$\beta = -.71$, 95\% CI [-1.10, -.32]), though its impact was moderate by comparison. These findings indicate that participants allocated significantly less of their cycling time to the target HR zone under both \textsc{Baseline} and \textsc{Random NPC}, relative to \textsc{Adaptive NPC}. Consequently, the \textsc{Adaptive NPC} condition most effectively helped participants maintain optimal HR across all target zones (\autoref{fig:results_hr_ratio}).

\subsection{Subjective Results}

\subsubsection{Borg Rating of Perceived Exertion (RPE)}

The intercept of the model, corresponding to the \textsc{Adaptive NPC} condition, was estimated to be 11.82 (95\% CI $[10.77, 12.88]$, $t(46) = 22.54$, $p < .001$). Neither the \textsc{Baseline} scenario nor the \textsc{Random NPC} scenario significantly affected perceived exertion. Specifically, the effect of \textsc{Baseline} was statistically non-significant and negative, with a small effect size (\( \beta = -.88 \), 95\% CI $[-2.08, .31]$, $t(46) = -1.49$, $p = .144$). Similarly, the effect of \textsc{Random NPC} was also statistically non-significant and negative, with a low effect size (\( \beta = -.06 \), 95\% CI $[-1.25, 1.14]$, $t(46) = -.10$, $p = .921$). These results suggest neither the \textsc{Baseline} scenario nor the \textsc{Random NPC} significantly affected participants' perceived exertion as compared to the \textsc{Adaptive NPC} condition.

\subsubsection{Physical Activity Enjoyment Scale (PACES)}

Overall, the model's explanatory power was very weak (conditional $R^2 = 1.52 \times 10^{-3}$), with the fixed effects contributing minimally to the model's explanatory power (marginal $R^2 = 1.26 \times 10^{-4}$). The intercept of the model, corresponding to the \textsc{Adaptive NPC} condition, was estimated to be $4.45 \times 10^{-3}$ (95\% CI $[4.19 \times 10^{-3}, 4.72 \times 10^{-3}]$, $t(967) = 32.56$, $p < .001$). Neither the \textsc{Baseline} condition nor the \textsc{Random NPC} condition significantly affected participants' enjoyment of physical activity, as measured by the PACES scale.

Specifically, the effect of the \textsc{Baseline} condition was statistically non-significant and positive, with a minimal effect size (\( \beta = 3.85 \times 10^{-5} \), 95\% CI $[-3.37 \times 10^{-4}, 4.13 \times 10^{-4}]$, $t(967) = .20$, $p = .840$). Similarly, the effect of the \textsc{Random NPC} condition was statistically non-significant and negative, also with a small effect size (\( \beta = -2.80 \times 10^{-5} \), 95\% CI $[-4.03 \times 10^{-4}, 3.47 \times 10^{-4}]$, $t(967) = -.15$, $p = .884$). These findings suggest neither the \textsc{Baseline} condition nor the \textsc{Random NPC} condition compared to the \textsc{Adaptive NPC} condition significantly influenced participants' enjoyment of physical activity, as assessed by the PACES scale.

\subsubsection{Intrinsic Motivation Inventory (IMI)}

The model's total explanatory power was moderate ($R^2_{\text{conditional}} = .13$), with the part related to the fixed effects alone (marginal $R^2$) being $3.27 \times 10^{-3}$. Within this model, the effect of \textsc{Baseline} was statistically non-significant and negative, with a small effect size (\( \beta = -5.05 \times 10^{-5} \), 95\% CI $[-1.16 \times 10^{-4}, 1.53 \times 10^{-5}]$, $t(1525) = -1.51$, $p = .132$; Std. \( \beta = -.09 \), 95\% CI $[-.20, .03]$).  However, the effect of \textsc{Random NPC} was statistically significant and negative, with a small to moderate effect size (\( \beta = -7.96 \times 10^{-5} \), 95\% CI $[-1.45 \times 10^{-4}, -1.38 \times 10^{-5}]$, $t(1525) = -2.37$, $p = .018$; Std. \( \beta = -.14 \), 95\% CI $[-.25, -.02]$). The results indicate participants showed no significant change in motivation when transitioning from the \textsc{Baseline} to the \textsc{Adaptive NPC} condition ($p = .132$). However, participants' motivation significantly decreased when transitioning from the \textsc{Adaptive NPC} condition to the \textsc{Random NPC} condition ($p = .018$). These findings suggest an adaptive visualization designed to support an optimal HR ratio increased intrinsic motivation compared to a random visualization. 

\subsection{Summary}
Below, we recap and relate the key findings to our three research questions.

\vspace{-2mm}\subsubsection{RQ1: Real-Time Adaptations Increase Users’ Capacity to Maintain an Optimal HR}
Compared to the \textsc{Adaptive NPC} condition, participants in \textsc{Baseline} had a slightly lower normalized HR, while those in \textsc{Random NPC} exhibited the lowest normalized HR overall (see \autoref{fig:results_hr}). This suggests that \textsc{Random NPC} exerted the most substantial downward effect on HR, yet neither \textsc{Baseline} nor \textsc{Random NPC} significantly altered subjective effort, according to the Borg RPE scale. Despite \textsc{Adaptive NPC} producing the highest normalized HR, its presence did not inflate participants’ perceived exertion relative to the other conditions. These observations imply that real-time adaptive cues (like those in \textsc{Adaptive NPC}) help users maintain more precise HR levels \emph{without} increasing subjective fatigue, underscoring the potential of personalized adaptation for accurate training.

\vspace{-2mm}\subsubsection{RQ2: Adaptive Visualizations Support the User in Optimal Cardio Levels}

Participants in the \textsc{Baseline} condition spent significantly less time in the target HR zone than those in \textsc{Adaptive NPC} (\autoref{fig:results_hr_ratio}), and the same was true for \textsc{Random NPC}. In other words, \textsc{Adaptive NPC} emerged as the most effective method for helping users sustain an optimal HR across multiple target zones, aligning with our earlier evidence that real-time adaptations improve heart rate regulation. Notably, neither \textsc{Baseline} nor \textsc{Random NPC} conditions altered perceived exertion from the user’s perspective, suggesting that adaptive mechanisms can enhance workout precision without increasing subjective effort. These findings imply that real-time adaptive visuals may be a valuable tool for recreational athletes and professionals, allowing them to train with greater accuracy and physiologically-tailored exercise intensities.

\subsubsection{RQ3: Adaptive Visualizations Not Necessarily Support Motivation and Enjoyable Physical Exertion}

As measured by the PACES scale, neither the \textsc{Baseline} condition nor the \textsc{Random NPC} condition significantly altered participants’ enjoyment of physical activity compared to the \textsc{Adaptive NPC} condition. According to the IMI, transitioning from \textsc{Baseline} to \textsc{Adaptive NPC} did not yield a significant change in motivation ($p=.132$). However, participants displayed a significant \emph{decrease} in motivation when moving from \textsc{Adaptive NPC} to \textsc{Random NPC} ($p=.018$). This pattern suggests that adaptive visualizations, which help maintain an optimal HR ratio, may increase intrinsic motivation more effectively than merely adding a non-responsive NPC. 

Although gamification itself did not markedly influence motivation or enjoyment, the \emph{manner} in which the NPC is designed and behaves appears critical. While the study relied on quantitative measures, informal feedback suggested participants found the NPC engaging and intuitive. Future studies could investigate how NPC appearance and interaction variations further enhance intrinsic motivation and should incorporate structured qualitative interviews to validate these impressions.

\subsection{Limitations}

We relied on \citet{tanaka2001age} to estimate $HR_{\max}$ ($HR_{\max} = 208 - .7 \times \text{Age}$) in lieu of the more common $220 - \text{Age}$ equation \cite{cruz2014maximum}. This choice stemmed from evidence suggesting \citet{tanaka2001age} provides a slightly more accurate estimate for some populations. Nevertheless, alternative formulas, such as \citeauthor{karvonen1984physical}'s method \cite{karvonen1984physical}, may yield different zone thresholds. A logical extension of this work would be systematically comparing these baseline computations to evaluate any performance or usability trade-offs in exergame feedback.

Our study design also involved multiple varying conditions (e.g., NPC presence, visualization style, and feedback modality). Although our central comparison between Adaptive and Random NPCs effectively isolates the impact of physiological adaptation, since both conditions include an NPC,other interacting factors remain entangled. Additionally, the short-term duration of our study limits conclusions about long-term effects on motivation and adherence.

Another limitation lies in the cycling-specific nature of our visualizations and interaction tasks. While cycling integrates well with VR, other sports involve different biomechanical patterns and pacing constraints. This focus narrows the generalizability of our findings to other athletic contexts.

\section{General Discussion}

Our findings show that adaptive visualizations can positively influence users’ physiological responses and overall training effectiveness. 
Our findings demonstrate that real-time physiological adaptation can improve users’ ability to remain in their target heart rate zone without negatively impacting perceived effort or enjoyment. This supports the feasibility of incorporating bioadaptive mechanisms into immersive fitness platforms, where moment-to-moment regulation of exertion is beneficial, such as in home fitness, rehabilitation, or high-intensity interval training (HIIT) scenarios. Importantly, our design emphasizes engagement through embodiment, where physiological signals are not abstractly shown, but integrated directly into gameplay via NPC behavior. In this section, we revisit the key lessons from our research and discuss methodological considerations.

\subsection{Gamification Only Works in Combination With Adaptation}
A central insight from this study is the interdependence between gamification elements and adaptive features in sustaining user engagement and motivation. While a non-adaptive NPC may sometimes undermine motivation, pairing NPC or game-like feedback with real-time physiological adaptation can transform these elements into potent drivers of commitment. Specifically, the \textsc{Adaptive NPC} in our study was more effective at helping participants maintain their HR targets and feel intrinsically motivated than the \textsc{Random NPC}. This aligns partially with \citet{shaw2016competition}, who found that competing against an NPC boosted performance and motivation. However, in their work, the user’s previous session data controlled the NPC. In contrast, our \textsc{Random NPC} did not reflect any aspect of user performance, possibly leading to confusion and lower motivation.  

These observations indicate that gamification by itself may either increase or decrease engagement depending on how responsively it connects to the user’s actual training state. When integrated with real-time HR adaptation, NPCs become more than just animated companions: they offer feedback loops reinforcing effort and promoting skillful pacing. Future work should examine variations of NPC behavior, such as replay-based or cooperative models, to determine how different adaptive strategies affect both short-term performance and long-term motivation. Exergames can leverage gamification more consistently by refining these design choices to foster effective, engaging workouts.

\subsection{Adaptive Visualizations Are Well Suited to Support Interval Training.}
Rather than relying on numerical displays or passive indicators, our system renders physiological feedback as a spatial interaction, embodied by a virtual cycling companion. This design allows users to “read” their body’s performance through in-world cues, not external metrics. This approach aligns with broader HCI work on embodied interaction and tangible computing, where the boundary between system and body becomes more fluid and perceptually integrated.

Our findings indicate that the \textsc{Adaptive NPC} design not only improved participants’ heart rate maintenance but did so without elevating their perceived exertion. This outcome suggests that real-time adaptive feedback can be particularly beneficial for \emph{interval training} programs, which rely on frequent transitions between high-intensity and lower-intensity effort to enhance cardiovascular capacity. Real-time adaptation allows these visually guided systems to promptly indicate when users should adjust their intensity, facilitating rapid comprehension and more precise adherence to target HR zones (see \autoref{fig:optimalHRAdaptation}). In this way, an \textsc{Adaptive NPC} can outperform conventional static cues by reducing guesswork about pacing or intensity changes.

Additionally, layering in gamification elements, such as in-game rewards or progress markers, can help sustain user interest and motivation, making structured interval regimens more engaging and enjoyable. This synergy between adaptive feedback and gamification thus can enhance both the effectiveness and overall experience of interval training, ultimately leading to more consistent progress and better fitness outcomes.

\subsection{Why Motivation and Enjoyment Did Not Shift (Yet)}
The absence of significant differences in intrinsic motivation (IMI) or enjoyment (PACES) between conditions may reflect several factors. First, the session durations were brief (6 minutes per condition), which may not provide enough time for affective differences to emerge. Second, many participants were inexperienced with VR cycling, and cognitive resources may have been allocated to learning basic operation rather than assessing emotional responses. Lastly, while physiological adaptation improved exertion control, its novelty or emotional payoff might require repeated sessions or gamified progression to fully manifest. These limitations do not diminish the system’s contribution but instead point to its potential in longer-term deployment.

\subsection{Explore Long-Term Effects of Concepts.}
Our investigation adopted a phased approach. First, we surveyed a variety of Mixed Reality (MR) visualizations for cycling, identifying \textsc{NPC} and \textsc{GUI} as standout designs relative to a baseline. We then concentrated on the \textsc{NPC} concept in a second study that addressed three core questions about real-time adaptation’s impact on heart rate maintenance, perceived exertion, enjoyment, and motivation. While these findings underscore the short-term effectiveness of adaptive feedback in helping users achieve target HR zones, they reveal smaller gains for hedonic measures. This discrepancy is especially relevant if the goal is to keep amateur users engaged over longer periods, where enjoyment and motivation play a more decisive role.

Notably, although our \textsc{NPC} design enhanced physical performance, we did not observe substantial motivational benefits, contrasting with \citet{michael2020race}, who reported increased intrinsic motivation over a four-week interval. It is plausible that such motivational shifts become more pronounced with extended practice, suggesting the importance of long-term studies for a clearer picture. A third longitudinal step, tracking whether and how these adaptive exergame concepts sustain motivation over weeks or months, would further refine our understanding of how best to encourage regular engagement and real fitness gains.

\section{Future Work}
Several promising directions arise from our findings. A logical next step would involve systematically comparing different $HR_{\max}$ estimation methods (e.g., \citeauthor{tanaka2001age}, \citeauthor{karvonen1984physical}, and other formulas) to assess how these variations impact exergame performance, feedback precision, and user perception.

Given the limited motivational effects observed in the short term, future research should investigate how more sophisticated NPC designs, featuring lifelike behaviors, emotional expressions, and adaptive interactions, can enhance sustained user engagement. Longitudinal studies are especially important to evaluate how these adaptive elements influence motivation, adherence, and enjoyment over extended periods. Additionally, expanding beyond cycling to include diverse sports and movement dynamics will help assess the generalizability of adaptive exergame frameworks. 

Multi-user or competitive settings should also be explored, where real-time physiological feedback could enhance shared or socially driven training experiences. Ultimately, refining adaptive NPCs across varied contexts and timescales can support the development of more engaging, personalized, and effective VR fitness experiences.

\section{Conclusion}
This work contributes a modular framework for embedding biofeedback into immersive fitness training. These scenarios show how personalized biofeedback can become a game mechanic, not just an analytics overlay. By emphasizing real-time interpretability, our system makes physiological data accessible and actionable, especially for amateur users who often lack coaching or domain expertise.
We present a design space of heart rate-driven visualizations that includes gamification, environmental changes, and distortion-based feedback.
We introduce a novel interaction paradigm where physiological feedback is represented as the spatial position of an NPC, and provide the associated software stack for real-time adaptation.

\section{Open Science}
\label{osf}
We encourage readers to reproduce and extend our results and analysis methods. Therefore, our experimental setup collected datasets, and analysis scripts are openly available on the Open Science Framework (\url{https://osf.io/acpxy/}).

\begin{acks}
Francesco Chiossi was supported by the Austrian Science Fund (FWF) [I6682] as part of the project AIM: Multimodal Intent Communication of Autonomous Systems and  Project ID 251654672 TRR 161.
\end{acks}

\section*{AI Disclosure Statement}
\label{aidisclosure}
During the preparation of this work, the authors used OpenAI’s GPT-4 and Grammarly for grammar and style editing. All content was reviewed and edited by the authors, who take full responsibility for the final publication.

\bibliographystyle{ACM-Reference-Format}
\bibliography{bibliography}

\end{document}